\documentclass[acmsmall]{acmart}

\usepackage{xcolor} 
\usepackage[colorinlistoftodos]{todonotes}
\usepackage{verbatim}
\usepackage{booktabs}
\usepackage{xspace}
\usepackage{longtable}
\usepackage{makecell}
\usepackage{float}
\usepackage{wrapfig}
\newcommand{\jb}[1]{\textcolor{black}{#1 \xspace}}

\newcommand{\jc}[1]{\textcolor{black}{#1 \xspace}}

\AtBeginDocument{%
  \providecommand\BibTeX{{%
    \normalfont B\kern-0.5em{\scshape i\kern-0.25em b}\kern-0.8em\TeX}}}

\acmJournal{PACMHCI}
\acmYear{2025} \acmVolume{9} \acmNumber{7} \acmArticle{242} \acmMonth{11} \acmDOI{10.1145/3757423}

\begin{document}

\title[Should AI Mimic People? Understanding AI-Supported Writing Technology Among Black Users]{Should AI Mimic People? Understanding AI-Supported Writing Technology Among Black Users}

\author{Jeffrey Basoah}
\orcid{0009-0006-6159-3200}
\email{jeffkb28@uw.edu}
\affiliation{%
  \institution{University of Washington}
  \city{Seattle}
  \state{Washington}
  \country{USA}
  \postcode{98195}
}

\author{Jay L. Cunningham}
\orcid{0000-0003-2446-8022}
\email{jaylcham@uw.edu}
\affiliation{%
  \institution{University of Washington}
  \city{Seattle}
  \state{Washington}
  \country{USA}
}

\author{Erica Adams}
\orcid{0009-0004-7857-6189}
\email{evadams6@uw.edu}
\affiliation{%
  \institution{University of Washington}
  \city{Seattle}
  \state{Washington}
  \country{USA}
}

\author{Alisha Bose}
\orcid{0009-0001-7763-6416}
\email{abose04@uw.edu}
\affiliation{%
  \institution{University of Washington}
  \city{Seattle}
  \state{Washington}
  \country{USA}
}

\author{Aditi Jain}
\orcid{0009-0007-6963-7930}
\email{ajain04@uw.edu}
\affiliation{%
  \institution{University of Washington}
  \city{Seattle}
  \state{Washington}
  \country{USA}
}

\author{Kaustubh Yadav}
\orcid{0009-0000-4906-1427}
\email{kausty@uw.edu}
\affiliation{%
  \institution{University of Washington}
  \city{Seattle}
  \state{Washington}
  \country{USA}
}

\author{Zhengyang Yang}
\orcid{0009-0001-1931-7134}
\email{yanzy@uw.edu}
\affiliation{%
  \institution{University of Washington}
  \city{Seattle}
  \state{Washington}
  \country{USA}
}

\author{Katharina Reinecke}
\orcid{0000-0001-7897-9325}
\email{reinecke@cs.washington.edu}
\affiliation{%
  \institution{University of Washington}
  \city{Seattle}
  \state{Washington}
  \country{USA}
}

\author{Daniela Rosner}
\orcid{0000-0001-9448-914X}
\email{dkrosner@uw.edu}
\affiliation{%
  \institution{University of Washington}
  \city{Seattle}
  \state{Washington}
  \country{USA}
}
\renewcommand{\shortauthors}{Basoah et al.}

\begin{abstract}
AI-supported writing technologies (AISWT) that provide grammatical suggestions, autocomplete sentences, or generate and rewrite text are now a regular feature integrated into many people's workflows. However, little is known about how people perceive the suggestions these tools provide. In this paper, we investigate how Black American users perceive AISWT, motivated by prior findings in natural language processing that highlight how the underlying large language models can contain racial biases. Using interviews and observational user studies with 13 Black American users of AISWT, we found a strong tradeoff between the perceived benefits of using AISWT to enhance their writing style and  feeling like ``it wasn't built for us''. Specifically, participants reported AISWT's failure to recognize commonly used names and expressions in African American Vernacular English, experiencing its corrections as hurtful and alienating and fearing it might further minoritize their culture. We end with a reflection on the tension between AISWT that fail to include Black American culture and language, and AISWT that attempt to mimic it, with attention to accuracy, authenticity, and the production of social difference. 

\end{abstract}



\received{January 2024}
\received[revised]{October 2024}
\received[accepted]{February 2025}

\maketitle

\section{Introduction}
Advances in natural language processing (NLP) are increasingly influencing many people's lives by supporting their writing process. Basic word processors and other tools can now provide grammatical suggestions, autocomplete sentences, or even generate and rewrite text, as is the case for large language models (LLMs) like Open AI's ChatGPT. While these AI-supported writing technologies (AISWT) have been hailed for revolutionizing the future of work~\cite{eloundou_gpts_2023}, increasing productivity~\cite{cambon_early_2023}, and providing more equitable editing and writing help to a broad population
~\cite{chakrabarty_creativity_2024, lin_why_2023}, Computer-Supported Cooperative Work and Social Computing (CSCW) researchers have repeatedly pointed out potential issues with the underlying LLMs~\cite{adams_diversity_2020, de_hond_picture_2022, chan_limits_2021, jung_toward_2023, boonprakong_workshop_2023}. For example, datasets and models used to train LLMs have been found to be more consistent with the values of Western and White people than with other groups of people~\cite{santy_nlpositionality_2023}. Researchers have also discussed that databases and training data are often biased~\cite{gururangan_whose_2022} and that the syntactic focus of NLP means that context and the use of language are all too often ignored by artificial intelligence (AI)~\cite{schlesinger_lets_2018}. What this means in practice is that LLMs commonly contain racial biases, including against African American Vernacular English (AAVE). Toxicity detection tools, for instance, are more likely to label expressions in AAVE as toxic than the equivalent expression in Standard American English (SAE)~\cite{sap_risk_2019, wilson_columbia_1996, halevy_mitigating_2021}. LLMs have been found to struggle in both generating and interpreting AAVE and generally performing better in generating SAE~\cite{deas_evaluation_2023, groenwold_investigating_2020}. 
While a notable body of work has examined biases in LLMs, studies that examine how individuals~\cite{burkhard_student_2022, wenzel_designing_2024, ike_inequity_2022, palanica_you_2019,pal_user_2019, pyae_investigating_2018}, and particularly African American users~\cite{mengesha_i_2021, wenzel_can_2023, harrington_its_2022, brewer_envisioning_2023,cunningham_understanding_2024}, perceive their daily interactions with NLP tools have only just begun (see~\cite{noble_algorithms_2018,benjamin_race_2019}).

In this paper, we build on this growing body of CSCW and adjacent work by investigating how Black American users perceive AISWT. We pose the following research question: \textit{What are the expectations, apprehensions, and perceptions of Black American users regarding AI-supported writing technology?} To answer this question, we employ a qualitative approach to understand the perceptions (gathered through semi-constructed virtual interviews) and experiences (observed in real-life context of a remote user study) of Black American users in their interactions with AISWT. Specifically, we examined the prior impressions and reactions of 13 Black American users to using AISWT as part of word processing software (Google Docs) and LLM (ChatGPT). We chose to focus on expectations, apprehensions, and perceptions because they represent key aspects of a user's experience while engaging with technology~\cite{yi_what_2023,sackl_more_2017}. Examining Black American users' expectations allows us to identify the baseline experience they anticipate when interacting with AISWT. 
Analyzing apprehensions sheds light on the barriers that deter Black American users from engaging with AISWT. 
Investigating perceptions enables us to uncover how Black American users understand and interpret AISWT.
By addressing expectations, apprehensions, and perceptions, we aim to gain insight into the process of designing technologies in ways that emphasize not only functionality but also access, including tradeoffs revealed through this broadened engagement. 

Our study reveals the impact of AISWT on Black American users and their linguistic and cultural expressions. The findings underscore a prevailing sentiment among participants of a notable absence of consideration for Black individuals and groups in the development of AISWT, largely due to AISWT's failure to recognize commonly used names and words within Black communities. Discomfort arises when AISWT attempts to replicate AAVE, with participants perceiving it as making unwarranted assumptions and casting doubt on the source of these assumptions. The study also sheds light on the perceived inefficiency of AISWT's editing features and the technology's potential impact on the perception of competence based on conformity to SAE. Despite these challenges, a substantial number of participants recognize the benefits of using AISWT to enhance their writing style and appear more professional, highlighting a mixed perspective on the technology's utility.

\subsection{Author Positionality}
In line with CSCW calls for radical care~\cite{karusala_future_2021}, it is important to illuminate our positionality as authors and the unique perspectives through which we interpret the data. The principal investigator is an Black American scholar born in the United States (U.S.) to immigrant parents from Ghana. The authors of this study collectively hail from diverse cultural backgrounds, encompassing the U.S., Europe and Asia, with several of us identifying as people of color. Several of the authors identify with the communities we are working with, and we each took extensive care to be respectful of those communities. We recognize that by engaging in user studies, we build on a legacy of social science inquiry characterized by exclusionary and sometimes extractive research practices that reanimate legacies of anti-Black racism within the academy (see~\cite{harney_undercommons_2013}). 
In this respect, our analysis owes much to Black feminist scholars of slavery and visual culture such as Saidiya Hartman and Christina Sharpe who write about chattel slavery and its afterlives. For Sharpe, drawing on Hartman, contemporary life is always unfolding in the literal and metaphoric ``wake'' of slavery's violences. This spatial, temporal, psychic, and technological tension then demands a kind of ``wake work'' -- or what Sharpe describes as ``a mode of inhabiting and rupturing the episteme with our known lived and un/imaginable lives.'' One major lesson from this work is the importance of reckoning with racial suffering without rehearsing and reproducing that same suffering. This tension recalls Hartman's question of archival analysis: ``How does one revisit the scene of subjection without replicating the grammar of violence?'' For us, this critical positioning prompts a deepened commitment to interrogating the very methods we take up to examine perceptions of AISWT.
Informed by the lead author's experience of negative feelings within these technologies, we recruited participants who were already familiar users of the technology under study. Rather than pose direct inquiries about the technology's benefits and risks, we gleaned insights from our discussions with participants, allowing their experiences to reveal their nuanced relationships to AISWT. In the pages that follow, we describe this process of examining AISWT with concern and care for our participants and their often conflicting experiences of use.
\section{Background}
\subsection{Racial Equity and Cultural Alignment in Large Language Models}
Our research concentrates on the biases in NLP, distinguishing it from other AI domains due to language's intrinsic vulnerability to biases and its significant societal impact. Language is fundamental to human interaction, and AI systems that process and generate language are crucial as they become integral to daily activities, where biases can deeply influence societal functions and individual perceptions~\cite{ram_conversational_2017}. NLP's significant role in shaping public opinion and perceptions can lead to the reinforcement of stereotypes and unfair treatment if not meticulously managed~\cite{bolukbasi_man_2016,jakesch_co-writing_2023}. Individuals' reliance on flawed heuristics when interacting with AI-generated text can result in deception, judgment errors, or the spread of misinformation~\cite{jakesch_human_2023}.

A substantial body of research has critically examined fairness and representation in AI and machine learning systems, uncovering pervasive biases that traverse race/ethnicity, culture, and language. This scholarship includes studies on cultural biases within AI technologies~\cite{scheuerman_datasets_2021}, algorithmic biases in data handling~\cite{scheuerman_how_2019}, and the systematic biases present in technologies like facial recognition, which often perpetuate societal norms and prejudices~\cite{scheuerman_how_2020}. Other work has considered the harmful effects of LLM outputs, with attention to the capacity for LLM to ``morally'' self-correct biased outputs \cite{ganguli_capacity_2023} and to the placement of blame, suggesting that users hold designers and developers responsible over the AI systems themselves \cite{lima_blaming_2023}. Strands of this research have illuminated gender biases in systems that fail to adequately recognize non-binary and transgender individuals~\cite{hamidi_gender_2018}. Collectively, these studies underscore the need for a comprehensive reevaluation of AI system development from dataset creation to deployment, advocating for practices that ensure more equitable and accurate technological outcomes. This holistic approach to understanding and mitigating biases in NLP and broader AI applications highlights the unique challenges and critical importance of addressing these issues in technology development and implementation. 

Emerging from the efforts of critical technology scholars such as Safiya Noble, Timnit Gebru, and Simone Browne~\cite{noble_algorithms_2018, browne_dark_2015, gebru_race_2020}, and advocacy groups like the Distributed AI Research Institute and the Algorithmic Justice League~\cite{institute_dair_2024,league_algorithmic_2024}, scholars have traced disparities in equity among minoritized groups to racial bias in particular~\cite{ogbonnaya-ogburu_critical_2020,boonprakong_workshop_2023}. This racial bias is not isolated to the datasets and algorithmic models; it is also embedded in society --- baked into everyday interactions, ideologies, and infrastructures~\cite{benjamin_race_2019,hettiachchi_investigating_2021}. When viewing fairness in this critical context, we are tasked with examining consequences of \jc{socially misaligned LLM based on race, culture, and linguistic variation}~\cite{kharchenko_how_2024, helm_diversity_2024}. Focusing on human values in AI development reorients attention beyond training datasets for LLM and the technologies that rely on them, toward the people who experience and are impacted by them.

Examining aspects of \jc {racial  equity, linguistic inclusion, and cultural sensitivity within LLM becomes particularly crucial to triangulating the system dynamics of AISWT that affirm diversity of human experiences whether through training data or development practices} \cite{deas_evaluation_2023,jung_toward_2023,groenwold_investigating_2020, prabhakaran_cultural_2022, ghosh_generative_2025}. These models have largely been trained on large quantities of internet data which emerge from various sources such as open-source repositories (e.g. Hugging Face), social media and online communities, Wikipedia, and books from digital libraries~\cite{flugge_perspectives_2021, lampinen_cscw_2022, brown_language_2020}. Consequently, these training datasets are often produced with text and logics of White Mainstream English (WME), underrepresenting AAVE and other minority language variations. Hence the inequitable outcomes which emerge from LLM datasets representing the positionality, views, and constructs of dominant language ideologies, which are more aligned with Western, White, cis-normative, and educated groups~\cite{santy_nlpositionality_2023, field_survey_2021, lampinen_cscw_2022}. Additionally, research has shown that training datasets can be further imbued with human biases as data annotators are often non-diverse and impart individual perspectives that denigrate and undervalue the significance of AAVE~\cite{deas_evaluation_2023, kapania_hunt_2023, denton_whose_2021}. While training corpora composed exclusively of AAVE is available, the sources remain under-resourced, outdated, and often fail to capture regional and intersectional variations~\cite{deas_evaluation_2023, farrington_corpus_2023, wassink_uneven_2022}. Subsequently, these corpora are less likely to be integrated into LLM datasets for commercial language technologies, with the exception of models like Latimer.AI aka ``The Black ChatGPT'', which uniquely elevates the experiences of Black and brown people as one of the few commercially available LLM addressing this linguistic variant. 

Our work builds on this existing analysis of Black users' impressions of NLP~\cite{cunningham_understanding_2024, wenzel_can_2023, brewer_envisioning_2023, harrington_its_2022, mengesha_i_2021} with a particular interest in AISWT. We examine the perceived effects of racial bias in NLP practices, how they permeate within AISWT experiences for Black AAVE speakers, and perspectives on how people are affected. Deas and colleagues suggest that ``\textit{more work is needed in order to develop LLM that can interact appropriately with those who use African American Language, a capability that is important as LLM are deployed in socially impactful contexts}''~\cite{deas_evaluation_2023}. In this paper, we therefore address this gap in the literature to better understand Black American users' perceptions of these tools.

\subsection[AAVE]{AAVE\footnote{Over the years, this English language variety has been referred to by various names, including African American Vernacular English (AAVE), African American Language (AAL), Black American English, and Ebonics\cite{rickford_african_1999}. For the context of this work, we will refer to it as AAVE \cite{rickford_african_1999}.} Linguistic Bias and Language Technologies}

Within the scholarship on AI and responsibility, several works highlight the linguistic bias that exists within language technologies and NLP systems broadly. In examining the prevalence and impacts of such biases and these systems, we fix unique attention on Black American speakers of AAVE. It is well documented that NLP systems exercise preferential treatment for users of SAE, leading to disparities in technology performance for non-standard minority variations \cite{blodgett_language_2020}. These disparities manifest in a myriad of forms which often convey fairness-related harms of allocation, quality-of-service, erasure, stereotyping, and mis-representation of minority language groups including Black AAVE speakers \cite{blodgett_responsible_2022, mengesha_i_2021, cunningham_understanding_2024, koenecke_racial_2020, harrington_its_2022}. These harms are largely associated with equity short-comings in NLP development including: bias in datasets \cite{meyer_artie_2020}, bias in automated speech recognition \cite{koenecke_racial_2020, moran_racial_2021}, bias in toxic language detection \cite{sap_annotators_2022}, bias in text generation \cite{deas_evaluation_2023, groenwold_investigating_2020}, and bias in language identification \cite{blodgett_racial_2017}. These previous works underscore the importance of cultivating cultural sensitivity within language technology \cite{mengesha_i_2021, wenzel_can_2023}, aiming for inclusivity across diverse linguistic varieties and avoiding the perpetuation discriminatory language ideology, which can have detrimental effects on minority groups. 

As awareness of bias in language technologies grows, there is a concurrent increase in efforts to further understand, mitigate, and prevent its harmful effects. These approaches have included guidelines for fairness and collaboration~\cite{blodgett_language_2020}, toolkits for bias detection and mitigation~\cite{bellamy_ai_2018}, development of fine-tuned models~\cite{groenwold_investigating_2020}, and inclusive data collection~\cite{dacon_evaluating_2022}. 
~
And while most discussions on understanding biases and inefficacies in NLP concentrate on system-level performance~\cite{halevy_mitigating_2021}, there remains room for empirical contributions that surface a deeper context of how people perceive and experience bias within certain classes of language technologies. The focus should shift towards reimagining language systems to be more inclusively group-centered and culturally responsive to the needs of marginalized communities currently underserved by these technologies.

The research in this paper aims to fill a crucial gap in understanding the perceptions of Black American users towards AISWT, specifically focusing on how racial biases in NLP affect these users. The study builds upon existing work that suggests LLM often fail to appropriately interact with AAVE, a gap highlighted by Deas et al.~\cite{deas_evaluation_2023} who argue for more research into LLM that can accurately understand and use AAVE in socially impactful contexts. The paper seeks to explore the deeper personal and communal impacts of these biases by focusing on the experiences and perceptions of Black American users, moving beyond system-level performance issues to address how racial biases in NLP technologies influence the daily interactions and societal integration of these tools.
\section{Methods}
\subsection{Participants}
We employed a snowball sampling approach to recruit participants, utilizing a screening survey on various platforms. The platforms included the principal investigator's Instagram and LinkedIn, large group chats (with over 100 members, which the principal investigator was apart of) on GroupMe, and departmental Slack channels. In the screening survey, we gathered responses (\textit{n=172}) inquiring about participant frequency of using AISWT, as well as their demographics, including gender, race, age, and level of educational attainment. The participant recruitment message sought to gauge interest in joining a 1-hour virtual session discussing and engaging with AISWT. Participants needed to be aged 18 or older and self-identified as African-American and U.S. citizens residing within the U.S.. When addressing African-American people, we refer to the diaspora of people of African descent, regardless of ethnicity~\cite{caldwell-colbert_chapter_2003}. We focused on U.S. citizens residing within the U.S. to help ensure that participants have a certain level of understanding of African-American culture. Additionally, the focus of the study is on the experience of African-Americans as the definition of the Black race can vary internationally. We sought participants that self identified as possessing basic digital literacy and having prior experience with AISWT, specifically inquiring about their experience with text editing features like spellcheck and grammar check, autocorrect, and generative AI like ChatGPT and chatbots. Among the survey respondents, 71 qualified for the study. Of those 71, 13 participants (see Table~\ref{tab:Table1}) were available and successfully completed both the interview and user study phases of the study, and received a \$50 USD voucher. Several participants used AISWT in educational settings for class exercises and experimentation with prompts. AutoCorrect, Grammarly, and predictive text features were widely used across participants for daily writing tasks, including emails, document creation, and academic-related content. Participants extend their use of AISWT into professional settings, incorporating them into work-related projects, communication, and note-taking during meetings. For a detailed exploration of the participants' overall engagement with AISWT, see Supplementary Material. 

\begin{table}
\centering
\caption{Overview of participants who completed the study. Participants were given the option to self-select their own pseudonym, allowing them to maintain agency over their identity. \jb{This design choice was inspired by Kesewaa's work, which explores the dynamics of power and privilege in the renaming of participants.~\cite{kesewaa_dankwa_all_2021}}}
\label{tab:Table1}
\begin{tabular}{@{}ccccc@{}}
    \toprule
Participant ID & Pseudonym      & Age Range         & Gender & Highest level of education obtained \\
    \midrule
P1             & Purple Lizzard & 18 - 25 years old & Woman  & Bachelor’s                          \\
P2             & N/A            & 26 - 34 years old & Woman  & Bachelor’s                          \\
P3             & N/A            & 18 - 25 years old & Man    & Associate's                         \\
P4             & N/A            & 18 - 25 years old & Woman  & Bachelor’s                          \\
P5             & N/A            & 26 - 34 years old & Man    & Bachelor’s                          \\
P6             & Black Tiger    & 26 - 34 years old & Man    & Master’s                            \\
P7             & Blue Bird      & 26 - 34 years old & Woman  & Bachelor’s                          \\
P8             & MamaAfrika     & 26 - 34 years old & Woman  & Bachelor’s                          \\
P9             & N/A            & 18 - 25 years old & Woman  & High School Diploma/ GED            \\
P10            & N/A            & 26 - 34 years old & Man    & Bachelor’s                          \\
P11            & N/A            & 18 - 25 years old & Man    & High School Diploma/ GED            \\
P12            & N/A            & 35 - 44 years old & Man    & Bachelor’s                          \\
P13            & N/A            & 18 - 25 years old & Woman  & Bachelor’s    
\\
\bottomrule                  
\end{tabular}
\end{table}
\subsection{Study Design}
\subsubsection{Semi-Structured Interviews. }To gain insights into participant perceptions of AISWT, we conducted one-on-one semi-structured virtual interviews, delving into their thoughts on AISWT and its alignment with their lived experiences. We initiated the interviews by assessing participants' foundational usage of AISWT, specifically inquiring about AI text generators like chatbots, smart text assistants, ChatGPT, as well as autocorrect, grammar, or spell-check features on platforms such as iPhone, Microsoft Word, or Google Docs. Following this, participants were prompted to articulate their understanding of Black American culture and their day-to-day experiences in their own words. This served as a primer to facilitate discussions on the alignment or disalignment of AISWT in subsequent questions. The conversation then shifted towards exploring participants' perceptions of the mentioned AISWT. Several inquiries focused on gauging the degree of cultural alignment, exploring the potential challenges or benefits introduced by these tools to the Black community, investigating the incorporation of Black perspectives in their development, and understanding the role of these technologies in addressing or exacerbating challenges within the community.

\begin{figure}
    \centering
    \includegraphics[width=1\linewidth]{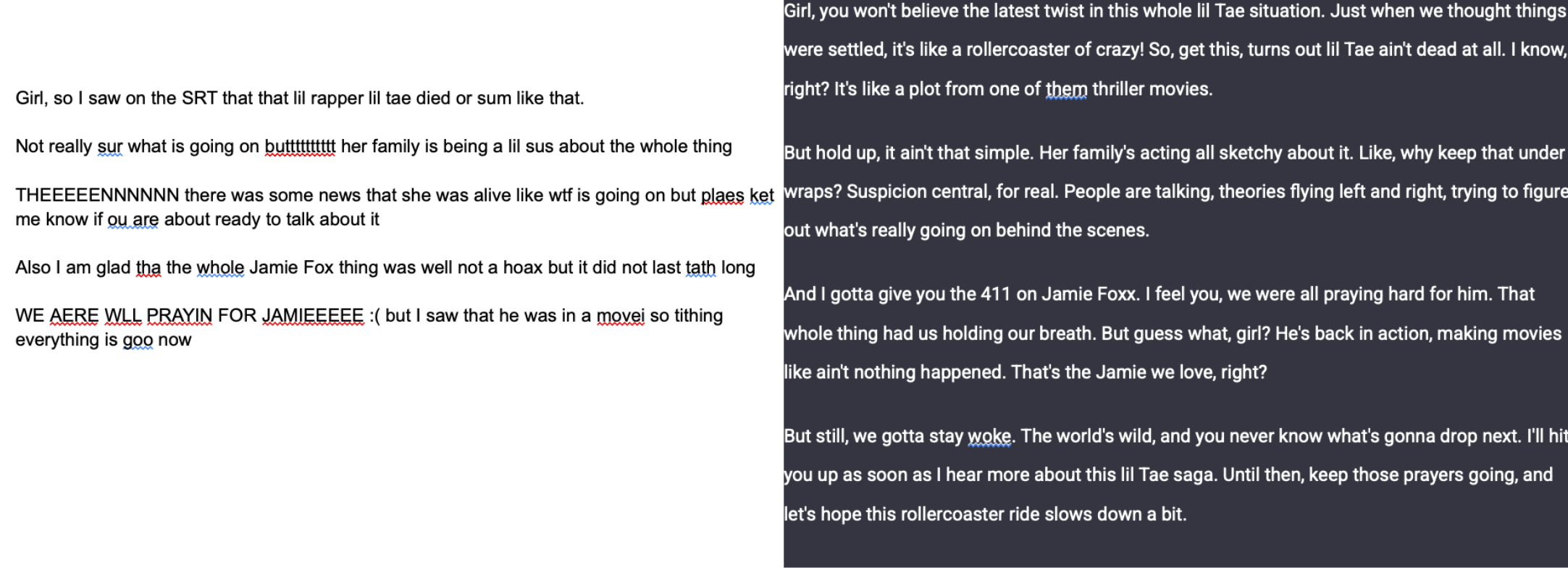}
    \caption{Comparison of Original Story Draft (Left) and AI-Generated Continuation (Right) during Remote Moderated User Observations. Participants engaged in an AISWT task where they first wrote a story in their natural vernacular, prompted by a casual writing prompt. The left side shows the participant’s original writing in their natural tone, while the right side illustrates ChatGPT's attempt to continue the story with consistent tone and vernacular, as per the participant's style.}
    \Description{The image shows a side-by-side comparison of two text sections. The left section is written in a document editor with a white background, while the right section is presented in a chat interface with a dark background. Left Side (Document Editor): This section displays a casual story written in African American Vernacular English (AAVE). The story is formatted in standard text, with some words highlighted in red due to spelling or grammar suggestions from the editor (Google Docs). The story discusses a rumor about a rapper's death and the author's reaction, using informal language and slang, such as ``lil sus''. The tone is casual and conversational, mimicking a text message format between friends. Right Side (Chat Interface): This section presents a continuation of the story using AI-generated text, keeping the tone and vernacular consistent with the original writing. The interface appears as a chat bubble on a dark background, with the same phrases and informal expressions as the original writing. The AI's attempt mimics the style and tone, continuing the casual conversation about the rapper's death and other related news. The purpose of this image is to illustrate the differences and similarities between a user-generated writing sample and an AI-generated continuation, both written in AAVE. The text is conversational, using informal slang and abbreviations}
\end{figure}

\subsubsection{Remote Moderated User Observations. } To understand participants' experience during AISWT interactions, we implemented a dual approach, utilizing remote moderated user observations and interviews. In our remote user study, we conducted observations by monitoring participants' interactions with the system through screen sharing sessions and analyzing their real-time feedback during the task. Observing the users' real time reactions (or apathy) to editing suggestions was crucial in providing meaningful context to the interview data~\cite{hillman_chapter_2015}.
While previous works such as Cunningham et al. (\cite{cunningham_understanding_2024}), Harrington et al. (\cite{harrington_its_2022}), and Mengesha et al. (\cite{mengesha_i_2021}) have utilized semi-structured interviews to gather insights from Black users regarding language technologies, our study extends this approach by incorporating real-time user observation. This allows us to build a more grounded and nuanced understanding of how users engage with these tools, offering insights that go beyond traditional interview methods. By merging interviews with usability studies, we aim to provide a more holistic view of user interactions, contributing to the broader scholarship on the topic and offering actionable insights for improving language technology design.

Interviews provided real-time clarification of responses, reducing the risk of misunderstandings. Our authors, whom identify as Black, took the lead in conducting interviews and direct observations to enhance researcher-participant connection and engagement, shaping the depth of responses \cite{davis_ethnic_2013}. 
Participants were provided with a writing prompt aimed at facilitating dialect elicitation. The rationale behind this approach was to offer participants an opportunity to express themselves in their natural vernacular without the influence of direct solicitation from us. By giving participants a prompt rather than specific instructions, we aimed to create a more organic and authentic environment for language expression. This method allowed participants to freely engage with the writing task, enabling us to capture a more genuine representation of their dialect and linguistic preferences. We provided participants with the following prompt: \textit{Pretend there is a time that you heard an interesting rumor/ gossip/ tea and you just had to text your bestie/ best friend. In at least 5 lines, we would like you to type out the story as if you were texting them now. Try to be as natural as possible in your writing, feel free to use slang or terms that you are most comfortable with. We are not here to test you but more so the technology that you are interacting with. Don’t worry about your grammar, spelling or anything of that sort. If you make a mistake, don't change or alter it}. The authors believed that recreating an environment wherein participants felt as though they were communicating with someone familiar would yield the most fruitful results for dialect elicitation as one finds themselves ``letting their hair down'' when communicating with a close friend or familial member. They were asked to write for 3 minutes in Google Docs in their “natural” style. We defined their natural style as their style of writing unencumbered by the pressures of meeting a writing standard. Next, we asked participants to utilize Zoom's screen sharing feature to share their screens and we discussed how Google Docs handled their text and what implications Google Docs' grammar and spelling suggestions may have on the participants' experience while using the word processor. The screen sharing was then switched to the interviewer, who transferred the participant's output and prompted ChatGPT to continue writing using the following input: \textit{Continue my story with an additional ten more sentences ensuring to keep my tone and vernacular consistent}. We then engaged in discussions with participants to explore their thoughts on ChatGPT's output. We focused on aspects such as their expectations of the output, the resemblance of the output to what they would produce themselves, and whether it met their anticipated results. We thought it best to have a user study of their experiences real time as it would help refresh participant memories of their experiences using AISWT outside of the study as well as allow them to add further context to interview responses previously given.

\paragraph{Ethical Review}
This study design was reviewed and approved by the University of Washington Institutional Review Board (IRB). The participants provided their written informed consent to participate in this study. Before participation, all individuals completed a screener survey where they were fully informed about the study's objectives, their participation roles, and their rights, including the option to withdraw at any time without consequences. This information was reiterated in their acceptance emails. Virtual written consent was secured at the start of the survey, ensuring participants' voluntary agreement before disclosing any information. To safeguard privacy, all data collected was anonymized, and any sensitive details were redacted. All electronic data were securely stored, accessible only to the research team.

\subsection{Analysis}
To understand our participants' perceptions and experiences with AISWT, our team performed a thematic analysis of the gathered interview data \cite{braun_using_2006}. We began our analysis process with inductive coding of two interviews. Six researchers coded two selected interviews to generate an initial list of codes. We then gathered all codes produced and began merging codes that were similar and used an affinity map to form larger coding groups. After developing our initial codebook, our research team analyzed the interviews in two stages. First, two researchers independently coded each interview once. Second, the two researchers met with the whole research team to discuss and iteratively adjust the codebook as necessary, accommodating emerging themes and discarding codes that were no longer applicable. This collaborative approach emphasized consistency and depth in our qualitative analysis. The memo book was instrumental in the final stage of analysis in which themes from data were derived using developed codes and relevant quotes. Following a process of community peer review ~\cite{liboiron_community_2018}, we invited participants to read and give feedback on our interpretations, analysis, and arguments, introducing a mode of mutual accountability into otherwise relatively established human-centered design practices. The codebook is included in Supplementary Materials to provide transparency regarding our coding scheme and definitions.  
\section{Findings}
The semi-structured virtual interviews provided valuable insights into our participants' perceptions, apprehensions, and expectations of AISWT. Concurrently, the remote usability study enabled us to delve deeper into their actual experiences with AISWT, offering a contextualized perspective to complement the information gathered during the interviews.

\jb{{\subsection{Significance of AAVE and the Limits of Mimicry}}}
In this section we delve into the significance of AAVE as a crucial mode of communication. Participants emphasized the cultural importance of AAVE and voiced their frustrations with AISWT's limitations in understanding and processing AAVE. This misinterpretation by AISWT lead to frequent autocorrections that distorted the intended message, causing inconvenience and dissatisfaction among users. This section underscores a crucial expectation among the Black American users in our study that AISWT should not only recognize but also accurately interpret and reflect AAVE to enhance communication rather than hinder it, highlighting a strong desire for technology that is culturally aware and capable of supporting diverse linguistic expressions.

\subsubsection{AAVE and its significance in communication}

We observed the cultural significance of AAVE for our participants and its impact on their communication expectations. Participants highlighted AAVE, occasionally labeled as slang in our conversations, as a vital mode of expression that they felt often went unrecognized due to AISWT's limitations in comprehension of the vernacular. Many participants shared their routine use of AAVE when communicating through text,  highlighting the challenges posed by AISWT when attempting to engage in AAVE-infused conversations.~\textit{``When I type in slang or try to use like slang terms, it will completely, like if I have my autocorrect on, it will just change everything and make it just like not make any sense at all''} states P7, a systems administrator based out of Washington, as she expressed frustration with autocorrect altering her slang-laden text to friends and family. Similarly P4, a Washington-state based graduate student in computer science, shared the inconvenience of AISWT not recognizing slang:

\begin{quote}
    \textit{``I guess maybe to some extent I found that you know when I'm texting, autocorrect will autocorrect you know some slang that I might use with my friends and that's a little inconvenient because I'd have to go back and change that or like go back and try to prevent that from happening [..] But I would say yeah, for things like autocorrect, it's not too great with like, more slang or like, informal texts that I use, and as a young Black person, like, I'm not going to text my friends in like a formal way. There's going to be some slang in there. So I do see it as an inconvenience when it does happen.'' (P4)}
\end{quote}

Despite facing these challenges, participants did not disengage from using AAVE. Instead, they perceived it as a deficiency in technology's robustness. P8, a portfolio manager of financial technology companies based out of Denver, echoed this sentiment as they describe their own experience interacting with AISWT: 

\begin{quote}
    \textit{``The way that we speak, especially colloquially, it's a lot more relaxed, a lot more informal, and I don't think autocorrect and these text correcting apps were built for that. And it doesn't really capture a lot of those like I said, like, `on fleek'. If I put fleek it probably gonna say `Did you mean flake?' and I'm like no, I meant fleek. It's not going to capture those little cultural and Black people isms. Shame on me for saying that but we have a lot of a lot of things that would not be captured there because it wasn't built for us.'' (P8)}
\end{quote}

Participants like P8 pointed out the limitations of these tools in capturing cultural nuances, expressing frustration when terms like `on fleek' (slang for `perfect' or `exactly right') are misunderstood. This concern resonated with the impressions of P2, a Virginia-based graduate student who relies on AISWT for writing support:

\begin{quote}
    \textit{``[It] doesn't keep up, doesn’t consider the evolving slang that we create or even it doesn't consider AAVE and how there's linguists, there's a whole study of linguistics on AAVE validating the fact that it's its own language, and the fact that [it] has not been considered like other languages is very disheartening.'' (P2)}
\end{quote}

P2 is highlighting the lack of consideration for evolving slang and AAVE by AISWT. Her statement aligns with participants expectation for technology to evolve in tandem with the diverse linguistic expressions inherent in AAVE. But it also acknowledges the existing validation of AAVE within linguistic fields, which AISWT fail to acknowledge or reflect. P13, a graduate student in informatics based out of Virginia, affirms this as she describes how her colleague feels uncomfortable using ChatGPT with AAVE as it often fails to recognize or acknowledge the nuances of AAVE, making using the platform not worthwhile:

\begin{quote}
    \textit{``They basically were saying they don't feel comfortable using ChatGPT with AAVE, even though that's how they text like they'd rather text and have something recognize what it's saying. But it doesn't always recognize Black or Black Vernacular English, or African American Vernacular English.'' (P13)}
\end{quote}

P13's observations highlight the importance of improving AISWT to be more inclusive and culturally sensitive. It also emphasizes the need for AI developers to consider the linguistic diversity and expressions of different cultural and linguistic groups. We see this sentiment echoed in Mengesha et al. (~\cite{mengesha_i_2021}) and Cunningham et al. (~\cite{cunningham_understanding_2024}), where participants highlighted the cultural insensitivity of speech recognition technologies to understand the nuances of AAVE. The inability of these language systems to accurately process AAVE further reinforces the notion that they are often developed with biases favoring dominant English varieties, leaving marginalized communities at a disadvantage. This mirrors findings from Koenecke et al. (~\cite{koenecke_racial_2020}),  Groenwold et al. (~\cite{groenwold_investigating_2020}), and Deas et al. (~\cite{deas_evaluation_2023}) where performance disparities in speech technology and text generation models revealed a consistent lack of support for diverse linguistic patterns like AAVE. These studies collectively emphasize the need for more inclusive and representative approaches in the design and development of language technologies.

\subsubsection{AISWT limited ability to imitate AAVE raises concerns about cultural understanding in technology}

Participants initially expected that AISWT would adapt to their communication style but found that it could only poorly imitate AAVE. For instance, like P1, a California-state based grad student studying Information Management, participants thought it \textit{``would be really interesting to see the model respond in that same language''}. \jb{However, P4, while interacting with the LLM, felt that although it put together a continuation of their story the output did not accurately reflect how she would naturally speak. She gave it an \textit{``A for effort,''} but not all participants were as forgiving.}

P13 expressed strong dissatisfaction with ChatGPT's attempts to mimic AAVE, believing that the technology should stick to answering questions rather than trying to continue the participants' statements. \jb{\textit{``I feel like ChatGPT has a space and it needs to hold its role, and that's like answering literally any other questions, but to continue what I'm saying like, it's almost disrespectful''} proclaims P13 as she reviewed the model’s output, \textit{``it's recognizing that slang exist, but it's not using it properly and like, for me, like that's, that's a little tricky.''} She found the model's use of slang to be problematic and felt that it was making unsuccessful attempts to draw from Black American culture.} She likened it to imitating language from old 70's movies featuring Black characters:

\begin{quote}
    \textit{``It's like it's drawing from like old 70's movies where Black people were in it. [..] I'm impressed that it generated this and just continued the story, but at the same time, I'm like, there's some serious issues that this offers where it's like one it's trying to mask and I guess imitate Black language and it's not doing it successfully.'' (P13)}
\end{quote}

P13's description appears akin to AI blackface, highlighting the perception of AISWT as an outsider invading their community. This underscores the sentiment that individuals from their communities are not the ones developing AISWT. These findings align with sentiments expressed in Cunningham et al. (\cite{cunningham_understanding_2024}), where participants attributed performance failures in language technologies to the lack of representation and inclusion of AAVE speakers in the design and development processes. This systemic exclusion leads to tools that fail to capture the linguistic nuances of AAVE, further widening disparities in user experiences. Additionally, this oversight can perpetuate harmful stereotypes, especially as NLP technologies play an increasingly significant role in shaping public opinion and perceptions, as discussed in Jakesch et al. (\cite{jakesch_co-writing_2023}). The biases embedded in language technologies can influence societal narratives, reinforcing pre-existing stereotypes and marginalizing underrepresented groups. 

Some participants felt protective of AAVE and believed it should remain within their communities rather than being diluted by widespread usage. P13 eloquently conveyed this sentiment, explaining that having the technology regurgitate their language did not make them feel comfortable: 

\begin{quote}
    \textit{``Having something else regurgitate that back to me does not feel great at all because like, that's what I use and that's what my community uses and I can't identify with ChatGPT like that. So the fact that that's being told back to me after I put my thing in there, it's like, oh, you're making a lot of assumptions right now [..] so it just doesn't make me feel comfortable.'' (P13)}
\end{quote}

Her discomfort arose from the sense that the technology was making unfounded assumptions about her language and culture, reinforcing the participants' perception that AISWT developers do not belong to their community. 

Participants expressed their desire not to be poorly imitated by AISWT but still yearned for genuine understanding. For instance, P6, a New York-based graduate student studying sustainability, shared their frustration with a tool like Quillbot, highlighting how it often misinterprets their use of correct terminology and proper language, resulting in a loss of comprehension:

\begin{quote}
    \textit{``That Quillbot, that's when I start to see I'm like, but I said it the right way or I use the proper terminology or you know, I said [..] things the right way, but it just completely changes it and it just doesn't understand it.'' (P6)}
 \end{quote}

A few participants expressed the idea that incorporating the ability to communicate in AAVE would be a beneficial feature for AISWT. P5, a graduate student in computer science based out of Texas, conveyed that enabling communication in AAVE would enhance the technology's language diversity, making it  \textit{``more inclusive for a broader range of users''}. P12, a project manager for an education policy organization in Washington, echoed this sentiment, suggesting that AISWT should  \textit{``mimic more like African American-based conversations''} to better embrace and include Black culture when asked on ways to improve the technology.

Participants grappled with the challenge of finding common ground between AISWT and AAVE. P13 suggested that AI should aim to understand the user's language and effectively communicate in AAVE when necessary. \textit{``AI needs to understand what they're saying, and be able to communicate that to that person and maybe it should be communicated in AAVE.''} suggests P13, also emphasizing the importance of avoiding stereotypes in the process, \textit{``But it also shouldn't be portraying a stereotype [..] So I agree with that, that there should be representation. It's just a matter of how it goes about it.''} She believed there should be representation in technology, but the manner in which it's achieved requires careful consideration. As highlighted in Deas et al. (\cite{deas_evaluation_2023}), there is a recurring challenge in the inability of LLMs to accurately generate and comprehend AAVE. \jb{This reflects broader concerns about control and inclusion in AI-mediated communication, where users feel alienated when AI tools fail to accommodate their linguistic identity~\cite{kadoma_role_2024}.}Our findings further complete this picture by illustrating the tangible effects of LLMs' limitations on Black American users, showcasing how these inaccuracies can alienate and frustrate members of marginalized groups.

\jb{\subsection{Apprehensions Emerge Around Misrepresentation and Cultural Erasure}}
The apprehensions expressed by our study participants are centered on concerns about cultural understanding, misrepresentation, and the potential for enforcing stereotypes. These fears contribute to hesitancy or outright avoidance of engaging with AISWT, particularly in contexts that relate to race and culture. These apprehensions highlight a broader distrust in AISWT ability to handle complex, culturally significant content with the sensitivity and depth it requires. The fear that these technologies could further entrench stereotypes and contribute to cultural erasure is a significant barrier to their acceptance and use among our Black American users. 

\subsubsection{Apprehensions that can lead to disengagement and avoidance of AISWT use}

Several participants expressed reservations about using AISWT in relation to their identity. When asked if there were specific examples of how AISWT could be exclusionary towards Black people, some were skeptical about whether the AI could genuinely comprehend issues related to social justice and the complexities of White supremacy. P2 highlights these concerns, particularly the fear of misinformation and the inability of the AI to address the intricate systems stemming from White supremacy, which perpetuates discrimination against marginalized communities:

\begin{quote}
    \textit{``So I think that can definitely go down a very, very slippery slope, especially if it spits out misinformation that we're already dealing with in media, social media, all those type of things we have to deal with, against the perception of people of color, especially Black people, White supremacist rhetoric, Lord knows. So that's my fear when I use it sometimes [..] and like the complex– it doesn't know how to break down the complex systems that stem from White supremacy, which is all like the things that marginalized people are discriminated against.'' (P2)}
\end{quote}

These apprehensions have led some individuals to completely avoid using the technology. For instance, P13 mentions that she \textit{``steers away from engaging in [AI] technology, explicitly relating to my race''}. When asked about the origins of this fear, P13 attributes it to a culmination of societal experiences. She emphasizes that topics like Black history often get overshadowed, and the historical implications and complexities are frequently overlooked, both in educational materials and potentially in technology. This under-representation contributes to their reluctance to engage with such technology:

\begin{quote}
    \textit{``Like I wouldn't use ChatGPT to generate like, a lesson plan about Black history [..] if you're making a lesson plan about history, and it only highlights like the, the highlights of an event, but it doesn't talk about like the historical implications of it that's problematic when it comes to Black people, because a lot of that, a lot of our struggle, is rooted in those implications. What are the results of the historical events that happen? Or were we there in those historical events? And so when you're talking about a topic like history, we get overlooked in the books, so chances are, we might get overlooked in technology.'' (P13)}
\end{quote}

P13's concerns raise a significant question: How can AI text technologies bridge the gaps in written content, which frequently present a biased perspective in favor of the victors rather than the victims? How can these technologies guarantee the delivery of a complete and balanced account of events to users? This inquiry builds upon the question posed by P1, ``\textit{who decides what is correct?}'' and extends it to ask, ``who decides what is true?'' Our participants’ reflections on the imbalanced representation of Black history and culture in language technologies are consistent with findings from other studies highlighting biases in favor of WME over AAVE \cite{deas_evaluation_2023, mengesha_i_2021,groenwold_investigating_2020, koenecke_racial_2020, agarwal_ai_2025}. Their skepticism and apprehension stem from lived experiences with performance disparities, where language technologies have historically struggled to accurately capture the communicative intricacies of AAVE. This inability to recognize or respect such linguistic nuance naturally extends to doubts about whether these technologies can fully engage with or represent the broader complexities of Black American culture. As a result, Black American users may justifiably question the inclusivity and cultural sensitivity of these systems.

\subsubsection{A collective concern about correctness and cultural diversity}

Many participants collectively voiced a shared concern about the divisive impact of AISWT on accepted and non-accepted language. This division becomes evident when participants see red squiggly lines under words they know are spelled correctly. Predating AI, popular word processing platforms, such as Microsoft Word and Grammarly, offer the appearance of a red squiggly line under misspelled words to indicates an error or how a text could be improved. Rather than viewing it as helpful,  our participants interpreted the line as a point of contention that stirs emotions of hurt in response to hate and sparks self-consciousness about their writing, leading to doubts about its validity and a pervasive feeling of exclusion. 
P1 aptly described this experience as if it's creating categorical distinctions, effectively segregating language into the correct and incorrect categories, stating \textit{``it's sort of creating, like those boxes I guess, like some sort of like, category that like, this is correct and this is incorrect.''} She further emphasizes the harm that arises when one form of language or speaking is deemed correct while others are labeled incorrect stating \textit{``it can be harmful when one language or one way of speaking is deemed as correct and the other way is deemed as incorrect''}. AISWT's corrective nature, as P9, an undergraduate student in computer science based out of Georgia, noted, exerts pressure for uniformity in communication, restricting any expressions that fall outside its predetermined boundaries. P9 stressed ``\textit{just because it doesn’t understand it doesn’t mean it’s wrong}”. This viewpoint is underscored by the figurative battle between features like autocorrect and spell check and AAVE, as P9 continued, ``\textit{the way it tries to change you or mold like your language to fit some something else, like a different community or different culture like no, we have our own, allow us to still be correct in our own}''. Participants expressed a genuine concern about the pressure for conformity imposed by AISWT. Participants, like P6 shared a fear of the impact the ensuing homogeny might have on the youth: 

\begin{quote}
    \textit {``I'm worried that like kids might use it and they might think this is the only way that you're supposed to speak, this is the only way that you're supposed to do certain things when it's not, you know, just AI doesn't understand culture.'' (P6)}
\end{quote}

P6 is underscoring that AI lacks an understanding of culture, and with the next generation increasingly turning to AI for answers, there is a collective fear that the technology's widespread integration could lead to the erasure of cultural diversity and community uniqueness. In an illustrative example, P6 recounted an interaction with ChatGPT where they asked it to create a dialogue between them and someone else. However, the generated text lacked the ethnic feel or the cultural nuances that P6 was accustomed to:

\begin{quote}
    \textit{``I was telling it to create dialogue between like me and somebody else and it was like, `Hey, dude, how are you doing?'' and I had to like try to go back and correct it a few times, just–just see what it was, you know, just how it will use how we use certain things and every time it was like, it didn't have that like ethnic feel.'' (P6)}
\end{quote}

ChatGPT was not able to replicate the dialogue that P6 was accustomed to, causing them to question \textit{``who programmed [it] to be this way'' and say, `Hey, dude' to say, you know, different things like that''} instead of language more representative of AAVE, which is more familiar and colloquial within his community. 

P1 raised concerns about the influences that AISWT corrections can have on users, largely due to the overconfidence they placed on AI's intelligence and access to vast online resources: 

\begin{quote}
    \textit{``You're reading the response, you also assume that AI is somewhat, you know, smarter than you are, it has access to the entire Internet and you only have access to your experiences of what you know, so it has to be correct.'' (P1)} 
\end{quote}

This blind optimism, paradoxically, leads participants to doubt their own knowledge and rely on AISWT adaptations, even when it provides out-of-context editing suggestions. This, as P1 suggests, \textit{``open up the risk of like, I guess like invalidating a type of language.''} P2 stated \textit{``AAVE is like it's a total other language in of itself''}, highlighting the uniqueness of AAVE, emphasizing that it's essentially a distinct language in itself.

Participants like P7 expressed the internalized constructs that form from the constant pressure to conform. This internalization leads to self-doubt and the feeling of speaking incorrectly or lacking proper communication skills: 

\begin{quote}
    \textit{``[It] makes us feel like well, when we speak it's like incorrect or we don't use proper English or we don't know how to talk properly or something.''} 
\end{quote}

While these experiences likely play out differently across activities (texting compared with document writing), participants like P1 and P7 emphasize the emotional toll of automated assistance through a rubric of correction that delegitimates alternative modes of communication. Despite these challenges, they also describe feeling compelled to accept the over-correction due to their limited choices, even though AISWT falls short of meeting their needs. The frustration, feelings of exclusion, self-consciousness, and doubt that participants expressed about their writing due to AAVE being marked as incorrect align with findings in several studies exploring Black users' experiences with language technologies. Research has consistently shown that language models and speech recognition technologies often fail to accurately process or validate AAVE, leading to negative emotional impacts. This mirrors the emotional responses seen in Wenzel et al. (\cite{wenzel_can_2023}), where participants reported lower self-esteem, self-consciousness, and diminished positive affect when interacting with speech technology, in Mengesha et al. (\cite{mengesha_i_2021}) where Black American participants reported that repeated misrecognition of their speech made them feel misunderstood or overlooked and in Harrington et al. (\cite{harrington_its_2022}) where participants felt inadequate due to voice assistants difficulty in comprehending their speech. These emotions suggest that Black users face substantial negative emotional impacts when engaging with both speech and text-based language technologies. What sets our participants' perspectives apart from previous studies is their concern for how future generations may increasingly rely on AI for information, potentially leading to the gradual erasure of Black American culture. These concerns are not unfounded, as research has shown that users often place undue trust in AI due to factors like overconfidence in the system's accuracy and the anthropomorphism of AI models~\cite{xinshuo_acceptance_2023, bi_i_2023, weidinger_ethical_2021, bubeck_sparks_2023}. But what happens when future generations are exposed to flawed imitations of AAVE or one-sided accounts of American history that fail to offer a holistic view? Our participants’ insights highlight the risk of misinformation, perpetuation of stereotypes, and the cultivation of self-doubt in the youth, presenting critical challenges for the preservation of cultural identity and the integrity of shared knowledge.


\subsubsection{AISWT serve a dual role in shaping professionalism and language gaps}

Contrary to the prevailing apprehension regarding the use of AISWT, particularly ChatGPT, a substantial number of participants highlighted the benefits of utilizing AISWT to enhance their writing style to appear more professional. These advantages were most apparent in situations where professionalism was essential, such as when communicating with colleagues via email or when establishing business relationships. When asked about AISWT benefits to Black communities, P3, a business analyst for a dairy company in New York, praised the resourcefulness of AISWT, as he saw them ``\textit{helping Black people to confidently communicate with folks all around the world that is correct and engaging}.'' P1 elaborated on how AISWT could be instrumental in creating opportunities for individuals who may struggle with effective communication, stating:

\begin{quote}
    \textit{``[..] it might be helpful to have some sort of predictive text, like some text generator that helps you curate an email that would maybe like, pass some sort of like HR, automated, like, email reviewing application, or even just like job application reviewing. So I think those things are helpful in terms of helping like, like people, [..] be able to communicate in a way that allows you to get your–your foot through the door, which is always the hardest part.'' (P1)}
 \end{quote}

Furthermore, P1 shared an example of how AISWT had aided her parents, for whom English was not their first language, in building relationships with business vendors:

\begin{quote}
    \textit{``I also think that autocorrect can be helpful when, for example, like, my parents use a lot of speech to text, like to respond to messages. So like, if they are pronouncing something like with an accent or something where it's not, it's not correctly written, it will appear with like the underline and say, you know, this is what you actually wanted to say, or it'll provide another option. So then they can be able to determine if like, oh, I should correct this.'' (P1)}
 \end{quote}
 
The suggestions offered by autocorrect and spellcheck serve as a double-edged sword. For non-native English speakers, AISWT seemed to bridge the gap in communication, while for those who speak English as their first language but in a different dialect, it could lead to misunderstandings. In a similar vein, AISWT also addressed internalized insecurities in line with their previously mentioned self-consciousness about their writing, stemming from editing suggestions that implied inadequacy, participants turned to AISWT to enhance their perceived intelligence. P8 shared her perspective, stating, \textit{``You want to sound a little bit more intelligent. So you're going to go to Grammarly [..] now you sound like an American person, or like a Caucasian person.''} When asked to elaborate, she clarified, 

\begin{quote}
    \textit{``[American and White] are not synonymous [..] but I think de facto the White experience is reflective of the nation. I think that's changing right with immigration and how different groups are rising in numbers [..] but yeah, they're not synonymous, but for some people they are, not for me.''}
\end{quote}
 
P8's statement encapsulates the sentiments shared by other participants regarding AISWT \textit{``Americanizing''} their writing. However, she clarified that she doesn't personally believe in this equivalence. These disrupted equivalences prompt the question of whether others expect AISWT to treat the White experience as representative of the nation and thus the American experience.

By pointing to changes, P8 and others bring a hopeful perspective to language technologies, differing from the challenges highlighted in prior studies \cite{brewer_envisioning_2023, pyae_investigating_2018, pal_user_2019, palanica_you_2019, wenzel_can_2023, wenzel_designing_2024, mengesha_i_2021, koenecke_racial_2020, harrington_its_2022, cunningham_understanding_2024, deas_evaluation_2023}. While past studies have focused on the limitations of speech technologies in handling non-English languages—often leading to uneven user experiences—our findings highlight how AISWT’s default alignment with WME can, in fact, assist certain demographics by improving their communication effectiveness. \jb{This dual role—both enabling and restricting linguistic expression—parallels findings in human-AI co-creativity research, where AI acts as a “second mind” in prewriting but also shapes users’ outputs in unintended ways~\cite{wan_it_2024}.}This points to the need for language technologies to be more dynamic, adapting to the user's linguistic background rather than forcing users to conform to a standardized mode of communication.

\jb{\subsection{Perceptions of Erasure and Inadequacy}}
The perceptions of our participants regarding the use of AISWT are heavily influenced by a sense of exclusion and a lack of cultural and linguistic sensitivity in the technology's development. Participants consistently noted that designers of AISWT seemed not to have Black communities in mind and failed to recognize and accommodate AAVE and the cultural nuances associated with it. Overall, they perceived AISWT as a tool that often fails to serve their needs adequately due to a lack of cultural and linguistic inclusivity. The technology's shortcomings not only hinder effective communication but also perpetuate feelings of exclusion and cultural erasure, underscoring the need for more thoughtful and engaged AI development practices.

\subsubsection{Exclusionary AISWT development and the need for inclusive, nuance-aware technology}

The study participants had a prevalent perception that the development of AISWT did not take Black individuals or groups into account, resulting in the exclusion of AAVE and Black culture. This consistent experience of words and names common within their community not being recognized by language technology leads participants to a collective conclusion that this technology simply, as P8 stated, \textit{``wasn't built for us''}. As succinctly articulated by P13 \textit{``[..] it wasn't created for Black people by Black people because [..] slang would be in there. Slang is not included at all. And so like, you know, it's frustrating [..].''} as they described whether they felt how Black people communicated through text was considered during AISWT's development. 

Many described the persistent issue of names common to them being flagged as incorrect or misspelled, while names seemingly associated with White individuals rarely faced similar flags. When discussing potential enhancements for AISWT, P9 shared the experience of encountering names prominent in Black culture flagged with a squiggly line as an error. She bluntly expressed the sentiment that ``\textit{these features were created just for the White man. Honestly, they didn't take into account Black culture as a whole.}'' P5 shares a similar opinion while discussing whether the technology had Black people in mind during development:

\begin{quote}
    \textit{``Probably not. When I type my name, my middle name, it will tell me my middle name is incorrect or that it's some kind of spelling mistake. So in that sense, no.'' (P5)}
\end{quote}

P5 is describing his own experience with his name as suggesting an exclusionary character built into word processing software like Microsoft Word. As if in response to this suggestion, P7 posits that this lack of consideration was intentional, driven by the belief that Black individuals ``\textit{aren’t going to really use them much}'' and so ``\textit{we don't really need to consider that there are other slang terms and things}'', resulting in product user experiences aligned with standards of dominant English speakers, who are often White. P10, a daily user of AISWT based out of Virginia, summarizes a pervasive sentiment:

\begin{quote}
    \textit{``I think we have to take a step back here and truly look at who it's been designed for. I don't [think] it is designed for [..] the Black community, because yeah, this doesn't tailor to us in any way. Because we weren't the target audience'' (P10)}
\end{quote}

P2 expressed their frustration, stating \textit{``[..] I hate it when Word, or like any other one of those word tracking things, so like Grammarly, they don't recognize, like, certain things like dialects and stuff like that, or like slang. And I'm just like, no, that's how you spell it.''} She goes on to explain how even names, such as their own or those of other Black individuals, are often marked as incorrect, intensifying their sense of self-consciousness: 

\begin{quote}
    \textit{``My name is spelt this way or just other–other Black people's names or just like anything that's not White. Like it just–I hate that red squiggly thing that comes under it just feels so.. I hate it. I hate it, absolutely hate it [..] I feel self-conscious about using that word because like it's squiggle has like the squiggly under it or it's like trying to correct it.'' (P2)} 
\end{quote}

This continued experience might result in a sense of otherness, continually experiencing the feeling of their name being singled out as incorrect, even though they are fully aware that it is not. P1, a graduate student based out of Virginia, recalls a conversation with a classmate who similarly encountered underlined words related to Black culture, which invoked a sense of \textit{``hate''} towards AISWT: 

\begin{quote}
    \textit{``[when she] sees like a name underlined or something underlined on her screen, it invokes some sort of emotion of like [..] hate [..] when that happens when it's really just someone's last name, may not be a very common last name.'' (P1)} 
\end{quote}

P1 is observing how having to constantly wrestle with these emotions and thoughts while using such technology can be taxing. Further reflecting on this experience, she described it as \textit{``exclusionary''} and pondered the same questions that prompted this study:

 \begin{quote} 
    \textit{``[..] some words that are used in the way that like, like language that's used by Black people may be considered spelled like incorrectly [..] because it's like, it's incorrect to who? And like, you know, it kind of brings that question of like, what is formal? Like, what is correct? What is to be considered a correct way of speaking? So yeah, no, I definitely think that there, there are aspects that are very much exclusionary.'' (P1)}
\end{quote}

This individual interpreted questions like ``\textit{what is formal?}'' and ``\textit{who decides what is incorrect?}'' as highlighting the inherent inequality in text correction features. She acknowledged the dual nature of the feature, as it seemed to constantly make judgments about what is permissible and what is not with every word.

Our participants’ sentiments align with existing research showing that LLMs tend to reflect the values and perspectives of dominant groups, particularly those aligned with Western, White, cis-normative, and educated demographics \cite{santy_nlpositionality_2023, field_survey_2021, lampinen_cscw_2022}. These models often reinforce dominant language ideologies, privileging SAE users while marginalizing others, as highlighted in Blodgett et al. (\cite{blodgett_language_2020}). What our participants are articulating is the impact of these biases, where they feel the exclusion and frustration of engaging with technologies that inherently favor demographics different from their own. 

Participants discerned a significant gap in AISWT's understanding of the nuances within Black culture, which some attributed to the absence of designers and developers from their community. Following the receipt of racist Trump rhetoric shortly after the introduction of a chatbot to his work chat, P12 attributed the unpleasant experience to developers who were not from his community. He expressed the belief that \textit{``it lets me know that like a racist White guy, a team of racist White people, were front and center on developing the chat.''}. Some others believe, as articulated by P4, that the technology is, \textit{``mainly built by like, White, or Asian men''}, contributes to feelings of exclusion, and they perceive any efforts to incorporate the Black community as a mere afterthought. 

Participants view any potential consideration for the Black community as an attempt to earn public approval rather than a sincere commitment to inclusivity. As exemplified by P4:
 \begin{quote}
    \textit{``You know, thinking about BIPOC people is kind of like an afterthought. Like, let's make sure that this technology works. And it'll be, you know, useful to people. And then maybe there might be some kind of monetary incentive down the line. And then it's like, oh, yeah, but then we also have to make sure that, you know, to prevent uproar, you know, that we put something in place to keep you know hate, the model from spewing hate'' (P4)}
\end{quote}
 
Participants like P4 observed that much of the current technology's inclusivity has stemmed from public outcry or a means to earn favor or to garner goodwill. They saw this continuous cycle of having to outcry before any change is made as leading to doubt if any inclusion is genuine or just an effort to maximize perceptions of social responsibility. 

Other participants shared insights on the potential for inclusivity by involving Black communities in the development process. When asked about opportunities to enhance AISWT for greater inclusivity of Black culture, P2 portrayed the relationship between developers and the Black community as a collaboration, highlighting the mutual benefits derived from the exchange of information. In this perspective, it is not a one-sided extraction of information but rather a partnership aimed at enhancing the product: 

\begin{quote}
    \textit{``It's a collaboration, it's not like a ‘I'm watching you to get this information extracted from you and never tell you about it’ it's more or less like now like we are exchanging information with each other so we can make this product better.'' (P2)}
\end{quote}

P13 is highlighting the importance of having individuals from their own community involved in the development process, expressing comfort in interacting with technology created by consultants or creators who share their cultural background. This approach, she suggested, bridges the perceptual gap created by the dominance of White and Asian architects within the technology's development:

\begin{quote}
    \textit{``[..] I would feel comfortable interacting with it, if it came from consultants, or a person who created it, who's part of my community, because I at least know that it's coming from the paradigm of the creator who looks like me.'' (P13)}
\end{quote}

This perspective emphasizes the importance of cultural representation within the creators and consultants responsible for designing AI and other technological tools. The involvement of creators from the same cultural community can be seen as a form of transparency, as it provides a clear link between the user's culture and the technology's development. This connection can strengthen cognitive trust, making users more receptive to AI and other technological solutions \cite{glikson_human_2020}. Our participants’ perspectives are strongly echoed in other studies that advocate for community involvement in the development of language technologies \cite{mengesha_i_2021, harrington_its_2022, blodgett_responsible_2022}. Engaging communities directly in the design and development process offers a pathway to addressing the cultural and linguistic gaps that currently exist in these technologies. 

Participants observed that AISWT falls short in capturing the intricacies of language, often favoring a ``one-size-fits-all'' approach. P4 underscores this by stating, \textit{``I don't think it really takes into account like different dialects of speaking, different ways of texting.”} They highlight that features like spellcheck and autocorrect seem to lack consideration for the linguistic variations of AAVE speakers. Participants shared instances where attempts have been made to account for the diverse nuances of language. For instance, P10 recalls working on a U.S. Census project that accommodated different French dialects, such as Canadian and Haitian. Reflecting on this experience, he expresses a sense of contrast, noting that AISWT lacks the same level of care and inclusivity in accommodating various modes of communication. P10 and P11, a college senior majoring in computer science in Virginia, characterized newly unboxed phones as being \textit{``more tailored to what a White person would say than what a Black person would say.''} They highlighted that these phones integrated with AISWT, don't initially \textit{``recognize everyone's slang''} upon first use, pointing at AISWT failing to embrace the linguistic diversity that Black individuals bring to the table.

The additional effort our participants discuss having to exert to interact with language technology is mirrored in Cunningham et al. (\cite{cunningham_understanding_2024}), where participants described the ``invisible labor'' of adapting their speech patterns to be understood by speech systems. This extra burden reflects the broader challenges faced by marginalized communities when engaging with technologies that fail to accommodate their linguistic norms. Similarly, Harrington et al. (\cite{harrington_its_2022}) observed that participants struggled to phrase their health queries in ways that voice assistants could comprehend, leading to feelings of frustration and inadequacy. This combination of extra labor and feelings of inadequacy raises important questions about whether the benefits of these technologies truly outweigh the drawbacks. 

\subsubsection{Disruptions and inefficiencies lead to feelings of inadequacy}

Participants highlighted the at times inefficient nature of AISWT editing features such as autocorrect and spellcheck. These features, originally intended to streamline the writing process, often lead to stress and frustration, disrupting the fluidity of communication and creative expression. Users found themselves needing to backtrack and make corrections when the technology failed to recognize or accept their intended text. These interruptions were not only cumbersome but also time-consuming, compelling users to invest additional effort to rectify them. P13 expressed feeling stumped during their writing process as \textit{``when [I’m] typing that into [Microsoft] Word it says like this is incorrect and then corrects it to the thing and I'm like, this isn’t what I wanted to say''} causing P13 to have to go back and reiterate what they have already typed. P13 suggested a \textit{``writer's form or writer's mode versus like academic mode or something like that''} but quickly retracted noting that \textit{``even having modes like that, where you labeled something as writers and academic and academic is associated with standard accepted English, it's still exclusionary.''} 

Technology's focus on SAE can lead to a perception that those who don't conform to this standard are less competent or considered less intelligent. P8 illustrated how this exclusionary design can affect how individuals perceive themselves and how others perceive them, potentially impacting their self-esteem:

\begin{quote}
    \textit{``The hindrances that anyone who doesn't kind of adopt to this format or this language is kind of left out of the picture, right? Or people might look down on you or think that you're not as intelligent or you can't spell things correctly, so on so forth.'' (P8)}
\end{quote}

These psychological consequences can have lasting effects, as the fear of inadequacy in communication may lead the entire community to rely on AI due to a lack of self-confidence in effective communication. These apprehensions and their potential long-term impacts are explored in subsequent sections.

\subsubsection{Participants feel overpowered by AISWT’s intrusive corrections, eroding their sense of language autonomy and privacy}

Some participants described their relationship with AISWT as one-sided, where the technology seemed to exert more influence over them than they had over it. P13 expressed frustration with trying to make adjustments to autocorrect and spellcheck features on their phone, only to find that the adjustments didn't work as expected. She felt that the technology didn't adhere to the parameters she set and wished it would simply allow her to type as she naturally does:

\begin{quote}
    \textit{``It’s frustrating when you have to go back [..] and retext something, or like be like, just ignore it and have to keep going and then my phone doesn't recognize it, either and I'm like I thought I was training you so that we wouldn't have to go through this again. But it doesn't even like you know, adhere to the rules that are in parameters that I'm trying to set [..] the thing that it could at least do is just let me type how I type.'' (P13)}
\end{quote}

This sentiment was shared by several participants who felt that they were being controlled by the technology rather than the other way around. P2 described the experience as feeling like she were being forced to \textit{``uncode''} switch in their personal messages and emails, which was invasive and unwelcome. \textit{``This is my own little chat bubble, like go away. Like I didn't, I didn't ask for you to correct me on something that I feel like I know is right, because like, again, like me and my community like we created this.''} states P2, as she emphasized the importance of having a space where she could communicate without external interference. The invasion of privacy by AISWT raises concerns about whether participants will have any private spaces left for their thoughts to be their own.

This finding parallels those in Harrington et al. (\cite{harrington_its_2022}), where participants felt compelled to adjust their natural speech patterns for voice assistants to understand their requests—a cognitively taxing process akin to code-switching. The mental strain of constantly altering speech in personal or intimate settings highlights the burden placed on users to adapt, rather than the technology adapting to them. Additionally, the constant back-and-forth adjustments participants had to make emphasizes the lack of a ``safe space'' for genuine self-expression while using language technologies. This further underscores how these tools can unintentionally suppress users' cultural and linguistic identities, creating barriers to authentic communication rather than fostering it.

\section{Discussion}

In this paper, we aimed to address the gap in the literature regarding Black American users' expectations, apprehensions, and perceptions of AISWT, focusing on the personal and communal impacts of biases within these technologies. Our approach moves beyond system-level performance issues to explore how racial biases in NLP technologies influence daily interactions and the broader societal integration of these tools.

While investigating user expectations, we found that our participants anticipated AISWT would accurately recognize, interpret, and reflect AAVE, a vital cultural mode of communication. However, participants experienced frustration with AISWT's limitations in processing AAVE, much like the findings in Mengesha et al. (\cite{mengesha_i_2021}) and Cunningham et al. (\cite{cunningham_understanding_2024}). The inability of AISWT to properly accommodate AAVE led to distorted auto-corrections, creating inconvenience and dissatisfaction among our participants. Participants expressed a strong desire for culturally aware technologies that supported diverse linguistic expressions. They also highlighted dissatisfaction with AISWT's unsuccessful attempts to represent Black American culture, which felt like stereotyping. This led participants to question the representation of Black developers in the design process, echoing sentiments from Cunningham et al. (\cite{cunningham_understanding_2024}) regarding the exclusion of AAVE speakers from the development of speech technologies. Interestingly, while many participants were concerned about the superficial portrayal of Black culture, some expressed a desire for AISWT to genuinely understand and authentically use AAVE, without falling into stereotypes. This nuanced perspective underscores the need for both cultural sensitivity and accurate representation in future iterations of language technologies, which has been underscored by existing research on AI and LLMs' values alignment~\cite{kharchenko_how_2024, helm_diversity_2024}.

As we explored user apprehensions about engaging with AISWT in relation to their identity, several participants expressed that AISWT often overlooks the nuances of communication within Black American communities. This led to a strong sense of exclusion, particularly concerning AAVE and its relationship to Black culture. Participants highlighted concerns around cultural misrepresentation and the risk of reinforcing stereotypes, reflecting a broader distrust in AISWT's ability to handle culturally sensitive topics. A recurring fear was the potential erasure of cultural diversity and community uniqueness, as participants saw AAVE and aspects of Black culture frequently marked as ``incorrect'' by AISWT. With the increased reliance on AI-driven technologies, users felt a growing pressure to conform to AISWT's adaptations, which, over time, led to self-doubt and diminished self-confidence—similar to the experiences reported by participants in Wenzel et al. (\cite{wenzel_can_2023}). Despite these frustrations, many users continued using AISWT due to a lack of better alternatives, reflecting a conflict between necessity and the emotional toll of such interactions. Interestingly, despite their apprehensions, participants also recognized some benefits, particularly in enhancing their writing style for more professional settings. This suggests that while AISWT can lead to alienation, it can also offer practical advantages, highlighting the complexity of its role in users' lives. This duality emphasizes the importance of addressing biases while also maximizing the helpful aspects of these technologies.

Our participant's perceptions of AISWT reflected a shared sense of exclusion and frustration, largely due to the lack of cultural and linguistic sensitivity in these technologies. Participants noted that AISWT often failed to recognize words and names commonly used within Black communities, leading many to conclude that these tools were ``not built for us'' and were not designed with the Black community in mind. This failure to accommodate AAVE and the cultural nuances tied to it underscored the inadequacies in how these technologies serve marginalized users. The experience of seeing red squiggly lines under AAVE words was not viewed as a suggestion for improvement but as a reminder of exclusion, eliciting feelings of self-consciousness and frustration. Our participants' sentiments align with research showing that LLMs often reflect the values and perspectives of dominant groups—those aligned with Western, White, cis-normative, and educated demographics \cite{santy_nlpositionality_2023, field_survey_2021, lampinen_cscw_2022, blodgett_language_2020}. These models tend to reinforce dominant language ideologies, privileging SAE users while marginalizing others. The frustrations expressed by our participants highlight the real-world impacts of these biases, illustrating how performance disparities in favor of Western, White norms further alienate underrepresented communities in their interactions with AISWT and language technologies as a whole. While some participants were skeptical of efforts to address the lack of inclusivity in language technology, others envisioned the potential for collaborative efforts between Black communities and developers. They emphasized the importance of having community members involved in the development process, echoing findings from other studies advocating for community engagement in language technology design \cite{mengesha_i_2021, harrington_its_2022, blodgett_responsible_2022}. Engaging directly with underrepresented communities offers a path toward addressing the cultural and linguistic gaps in these technologies, reducing alienation and fostering tools that better reflect users' identities and experiences. The additional effort our participants described that they must expend when interacting with AISWT is mirrored in Cunningham et al. (\cite{cunningham_understanding_2024}), where participants described the ``invisible labor'' of adapting their speech patterns to be understood by language systems. Similarly, Harrington et al. (\cite{harrington_its_2022}) reported participants' frustrations when their health-related queries were not properly understood by voice assistants, leading to feelings of inadequacy. This combination of extra labor and self-doubt raises critical questions about whether the benefits of these technologies truly outweigh the drawbacks, particularly for marginalized communities. While AISWT offers certain practical benefits, such as enhancing professional communication, it often comes at the cost of altering personal language patterns and suppressing cultural expression. These findings underscore the need for AISWT to evolve beyond merely accommodating dominant language norms and instead actively support linguistic diversity and inclusivity, allowing for more authentic and empowered user experiences.

In the following section, we reflect on three open questions in the broad area of AISWT and design for Black users: 
(1) How do we overcome the tradeoff between imitation and inclusion?
(2) How do we broaden the concept of trust to include authenticity?
(3) How do we bridge designing for and against social difference?

\subsection{How Do We Overcome the Tradeoff Between Imitation and Inclusion?}

\textbf{Participant Perspectives and Concerns. }One of the notable findings from our study was that participants preferred that the AISWT did not poorly imitate AAVE. While some participants questioned the intentions of mimicking and reproduction, others described hoping for greater comprehension and more accurate reproduction. The question of what it means for AISWT to ``accurately'' understand or replicate their language dialects, and AAVE in particular, prompted participants to describe feeling like they must adapt to the technology, rather than the technology adapting to users' needs. 
Some participants felt additionally protective of AAVE, expressing the belief that it should remain within Black communities to prevent dilution through widespread usage. 
 
\textbf{Broader Themes of Appropriation.} This grappling with accuracy 
recalls legacies of harmful appropriation within minoritized communities, and particularly the extraction of knowledge from Black and Native groups by research institutions~\cite[p.235]{tuck_chapter_2014}. \jb{To address such dangers, feminist and Indigenous studies scholars (~\cite{tuck_chapter_2014},~\cite{garcia_no_2020},~\cite{feminist_data_manifest-no_collective_feminist_2025}) have pointed to frameworks of refusal, which they define as ``attempts to place limits on conquest and the colonization of knowledge by marking what is off limits, what is not up for grabs or discussion, what is sacred, and what can’t be known.''} 
In our work, participants felt resistant to similar attempts at appropriating Black American linguistic and cultural knowledge.  

\textbf{Connections to Language Technology.} A connected apprehension has been outlined by Mengesha and colleagues, who in studying behavioral and psychological impacts on language technology errors, found that African Americans perceived the need to adapt themselves to be understood as a signal of being ``outside the group the technology was built for''~\cite{mengesha_i_2021}. 
The additional labor of adapting to SAE, coined as ``uncode switching'' by P2, introduces an unnecessary complexity for users, but also replicates a social barrier that minoritized communities often encounter when engaging with mainstream research institutions and dominant social groups. \jc{ \citet{cunningham_understanding_2024} similarly found that AAL speakers exude additional labor in language technology interactions that are often unaccounted for.} 
By replicating these challenges in their interactions with Black American users, AISWT further entrench those inequities, placing an additional burden on those users to adapt (``uncode'' switch) or refuse the tool.

\textbf{Proposed \jb{Direction.}}Rather than focus on minimizing risk or justifying human-centered design methods, our work draws attention to the complications that human-centered assessments of accuracy and user experience present. Recognizing the tradeoff between inclusion and imitation--or accuracy--as not an autonomous state involves treating the phenomenon as an emergent process shaped by the people, situation, and land it affects. It suggests CSCW scholars embrace a relational view of accuracy~\cite{littletree_centering_2020}, one that is situated in a social and material context and sensitive to the conditions of its development, practice, and performance. In this repositioning, our analysis calls for a careful consideration of what accountability practices (such as community peer review~\cite{liboiron_community_2018}) might be necessary to build into our user research tools as CSCW scholars. Just as P2 described the desired relationship between developers and Black communities as a collaboration, highlighting the mutual benefits derived from the exchange of information, we highlight the possibility of reckoning with troubling genealogies of conventional human-centered design assessments of imitation, accuracy, or communication. We propose a new \jb{direction}for assessing a technology's ability to adapt to and accommodate the diverse cultural nuances of its user base. This expands the criteria by which technology is evaluated, emphasizing not just its static performance but also its dynamic capabilities. It challenges developers and CSCW scholars to consider not only the current user but also the potential ways in which the technology may be used, and how it can be prepared to adapt accordingly. This approach shifts the focus from accepting out-of-the-box models to expecting adaptability, driving a more proactive and user-centered approach to development.


\subsection{How Can We Broaden the Concept of Trust to Include Authenticity?}
\textbf{Challenges with AAVE in AISWT.} A pervasive tension baked into AISWT concerns AAVE's complicated roots in both falling from and conditioning social difference. As many critical linguists and archival scholars such as Alicia Beckford Wassink and Marisa Feuntes \cite{fuentes_dispossessed_2016,wassink_historic_1999, wassink_sociophonetic_1999} have discussed, AAVE and adjacent languages exist in part due to the terrors of chattel slavery that underlie legacies of inequity and social difference for Black Americans today. But AAVE also exists as a forceful and nurturing influence on the lives of Black American individuals and communities in persistent and liberatory ways. In this dual status as both legacy of violence and nurturing potential, AAVE positions AISWT as playing a complicated role. AISWT may reinforce violent legacies of anti-Black racism by rehearsing stereotypes, extracting and appropriating content, or perpetuating creepy, uncomfortable, or devaluing engagement. Yet they may also support better attunement to the language grammars and performances that Black Americans engage every day. What it means to embolden or enliven AAVE without falling back on techniques that further entrench structural exclusion and harm requires asking deeper questions about the conditions by which people come to engage and ultimately trust a system. 

\textbf{Reframing Trust in AI.} As we know from prior work on trust and AI, the assessment of trust or trustworthiness often hinges on concerns for user compliance or “a predictor of user acceptance” \cite{glikson_human_2020, rai_explainable_2020, hancock_ai-mediated_2020,lee_understanding_2018}. Gilkson and Woolley’s \cite{glikson_human_2020} review of empirical research on human trust in AI, for example, identified a willingness to be vulnerable or take a meaningful risk as one definition of trust that translates across disciplines. From this perspective, one rooted in management science and social physiological precepts, trustworthiness becomes an enumerable predictor of action, an index for whether a user weighs the risk of acceptance as sufficiently worthwhile.


Our analysis of AAVE points to an alternative understanding of trust. Rather than consider risks and uncertainties, our participants drew attention to misaligned values and feelings of discomfort--a response best captured by questions of authenticity. What our participants felt or accepted as authentic involved more than a reading of trust as compliance: a conditioning of the user to believe or accept an interaction (answer, suggestion, or correction). Instead, it involved a particular concern for mutuality and transparency. It prompts questions, such as:  Where does an authentic engagement live within an AAVE simulation? And who or what is behind it? This reciprocal concern recalls a politics on consent discussed by Kinnee and colleagues \cite{kinnee_autospeculation_2023} wherein a researcher with a participant, just like a user with AI, must take care to check in and make space for connection as well as refusal. Our participants’ concerns for authenticity in AI then brings new readings of engagement back to conversations on algorithmic trust that highlight the need for assessing degrees of consentful and transparent interaction. 

\textbf{Future Directions for AISWT.} Our study suggests that future developments in AISWT, and AI in general, could focus on building a rapport with users before and during their interactions. The purpose of this rapport-building is to shift the experience from a ``one size fits all'' approach to a personalized interaction. This requires AI to be dynamic and flexible, adapting to individual users while respecting their consent and input. In doing so, users help shape a technology that meets their needs, rather than adjusting themselves to fit the limitations of the system—an issue many participants highlighted in their interactions with AISWT. As this connection develops, the user remains in control of what information they choose to share with the technology, shaping the experience they wish to have. By establishing this connection, the technology could address the sense of intrusiveness participants felt when receiving suggestions. 
Rather than users perceiving the technology as not made for them, the goal would be for it to adapt to them personally. This approach addresses one of the major challenges in AI: incorporating the unique cultural and community nuances of diverse users.

By shifting some of the responsibility to the user, AISWT can contribute to engaging cultural context. This involvement may increase the likelihood of a more personalized and culturally attuned experience. 
But it may also place an additional burden on users, requiring those who already face forms of cultural and racialized exclusion to do additional work to adapt an interface. Moreover, while AISWT's efforts to make users feel that the technology is designed with them in mind may bolster user trust, this does not necessarily mean that the firms creating these tools are held accountable for respecting user needs around privacy and control. Our findings suggest putting robust safeguards in place that assure consentful processes attend to the range of undue burdens and potential strategies that users may take up. In this sense, an emphasis on flexibility and transparency requires an equal attention to refusal: enabling users to shut down, disengage, or otherwise reject AI offerings in favor of non-use.  

\subsection{How Do We Bridge Designing for and against Social Difference?}

\textbf{Designing for Social Difference.} A final open question raised by our study involves what feminist Black Studies scholars such as Saidiya Hartman have referred to as the ``double bind''--the particular configuration of Blackness vis-à-vis a social order, and in this case the social order of an AISWT. Is it possible, or even preferable, to design for the liberatory experiences of the ``Black user'' when those categories operate as, and potentially reproduce, social difference? The idea of designing for a group of people who are historically and structurally differentiated may remediate existing harms or serve to reinforce and reproduce those gestures of differentiation. Technology may remake social differences just as it recognizes those differences. What it means to design for Black users is then tied up in a conversation on what takes priority: recognition or reimagining?

Our study suggests that the order of operations is less important than the coupling wherein recognition of social difference comes with a commitment to understanding social difference as never fully determining. Emerging apps like Latimer.AI point to this possibility. Dubbed as ``The Black ChatGPT,'' Latimer.AI is a LLM designed to provide a more accurate representation of the experiences, culture, and history of Black and Brown communities. In our study, we saw how frequently participants did not see themselves reflected in the technology, but also how often they did not expect to be seen. Those expectations reflect their expectations for who has designed the system as well as who manages and makes decisions that shape how that system affects their everyday lives. Changing the composition of a design team may change how particular features are implemented, but celebrating incremental changes as sufficient may also lend legitimacy to a wider sociotechnical infrastructure built to unevenly extract value from Black users.

\textbf{Reworking Systemic Foundations. }Black ChatGPT does not solve this double bind, but it helps expose how treating the double bind as a problem to “solve” or even resolve may be beside the point (see \cite{cunningham_grounds_2023}). As a set of design practices and performances, technological experiments like Black ChatGPT may instead challenge and rework the grounds on which systems are built—for example, how user activity gets treated as data, and who or what becomes the steward of that data (or those traces of activity) once it is created and shared. Within CSCW, Black ChatGPT can be seen as an artifactual outcome of social justice-oriented design, which attends to the ways that marginalized groups experience oppression and inequality within a society \cite{dombrowski_social_2016}. Dombrowski and colleagues assert that a social justice orientation in the design of technological artifacts can afford new practices, social habits, and ways of interacting that are informed by experiences and sensitivities of marginalized voices \cite{dombrowski_social_2016}. To this contention, we ask: What might AISWT look like if they were reimagined and informed by dynamics of Black American communities? 

\section{Limitations and Future Work}

The relatively youthful composition of our participants, with all but one under the age of 35, raises questions about the diversity of perspectives captured in our data. Given that many millennials and Gen Z individuals have grown up with language technology integrated into their lives, or have witnessed its increasing prevalence, our findings may lack the insights of those from a generation that encountered such technology in their adult years. Exploring the perspectives of this demographic could provide valuable insights into the influence, or lack thereof, of AISWT on their lives. 

The use of the snowball recruiting method in our study raises concerns about the diversity of our participant sample. This method, which relies on current participants to recruit future participants, may have contributed to the homogeneity observed in our study, where nearly all participants were under the age of 35 and the majority of our participants are highly educated. To address these limitations and broaden the diversity and educational heterogeneity of the perspectives represented in our dataset, we are committed to expanding our recruitment efforts beyond the lead author's network to better represent a wider range of perspectives, especially from groups that are currently underrepresented.


To enhance validity, we would examine these experiences across diverse contexts and activities, such as informal texting and formal document writing, recognizing that tool usage and user interactions with grammar correction features may vary significantly between these settings. For instance, while grammar correction tools in platforms like Google Docs are commonly used in professional or academic contexts, their relevance may be minimal in casual conversations, such as texting close friends, where language formality and error correction tend to be less prioritized. This broader approach would allow us to capture a more ecologically valid understanding of how and when these tools are actually utilized, providing richer insights into user behavior across both formal and informal language settings

Additionally, the voluntary nature of participation could have influenced the range of opinions captured, as those with strong views on the topic, either positive or negative, might have been more inclined to participate. This self-selection bias may have excluded more moderate or indifferent perspectives, potentially affecting the diversity of insights into the use and perception of AISWT.

Our choice to emphasize U.S. citizenship as a selection criterion to ensure participants' understanding of Black American culture has proven to be less robust upon reflection. Citizenship alone does not necessarily correlate with a deep understanding of the cultural nuances within the United States. This requirement has inadvertently excluded potential participants from the rich pool of the Black Diaspora, who may possess valuable insights. 

Upon reflection on our findings, our team engaged in a thorough discussion concerning the impact of our study design on the breadth of our discoveries. Our study was intentionally crafted to center around AISWT, which helped to focus the findings but also mitigates generalizability to other areas of AI. A more expansive inquiry into various AI tools could yield a more diverse set of responses in the future, particularly concerning the inclusion or exclusion of different aspects of Black culture. 

\section{Conclusion}

This paper explored the expectations, apprehensions, and perceptions of Black American users regarding AISWT, including word processors that provide grammatical suggestions and autocomplete sentences, and more advanced tools like ChatGPT that generate and rewrite text. By interviewing Black American participants (\textit{n=13}) and observing them interact with word processing software (Google Docs) and with LLMs (ChatGPT), we were able to learn about their past experiences with AISWT while also capturing their immediate reactions to receiving various suggestions from the AISWT. Our findings paint a picture of conflicting feelings. On the one hand, our participants were frequent users of AISWT and most found that these tools were helpful, such as to enhance their writing style. On the other hand, a majority of participants mentioned how AISWT's suggestions are inherently not designed for Black American users, especially because the tools usually highlight words common in AAVE as incorrect. Beyond a feeling of discomfort about the AISWT's corrections and writing suggestions, participants worried that ultimately, the Whiteness of these tools could eradicate their language and culture. The findings suggest a dilemma between living with software that ``is not designed for us'' and wanting it to become better at understanding Black American culture and language, the latter bringing up questions of authenticity and trust. Our participants' insights suggest a way forward: a technology that respects and adapts to diverse linguistic and cultural expressions, promotes language autonomy, and strives to understand, rather than merely imitate, the rich tapestry of human communication.

\section{Funding}

Funding for this study was provided through grants awarded by the Human Centered Design and Engineering program at the University of Washington. 

\begin{acks}
Any opinions expressed in this material are those of the authors and do not necessarily reflect the views of the University of Washington. We thank the reviewers at the Odegaard Writing and Research Center for their discussion and useful feedback. 
\end{acks}

\bibliographystyle{ACM-Reference-Format}
\bibliography{UBUEResearchNew}
\clearpage

\appendix

\section{Interview Protocol}
\subsection*{Introduction}
Script: Hello [Participant Name], thank you for taking the time out of your day to join us. Your participation is greatly appreciated by myself and the team. In this study, we are trying to understand what aspects of digital technology Black users find takes into account their lived experiences and highlight possible pitfalls of how current digital tech is designed that should be addressed. I am going to start by introducing myself and my partner and reviewing how this session will go. \\

My name is [Interviewer name], and I am a [UW class/program/etc.] and I’ll be serving as your interviewer today. I’m accompanied by my notetaker who will be taking notes for the interview. The interview will consist of a series of questions and prompts surrounding your perceptions and experiences using AI-supported text technology. We anticipate this interview portion being roughly 30 minutes and an observation of you using the technology to take about 15-20 minutes. There are no right or wrong answers to our inquiries we just ask that you share openly honestly and freely - we are here to learn and listen from you! What we talk about is confidential and will only be shared with members of the research team. You may choose to leave your camera on or off it is up to your discretion. \\

Before we continue are there any questions or concerns that you have for us at this time?\\

Great! Moving forward we would like to record this Zoom session today.
Do we have permission to record this interview? \textit{(In case of refusal, note-taker captures context manually)}.\\

Thank you! If you would like to conceal your identity before we begin recording we ask that you change your display name to a pseudonym of your choice (example: "Red Hippo").\\

Great! We will begin recording now.

\subsection*{Warm Up}
\begin{enumerate}
    \item Tell us a little bit about yourself. What do you do for a living?
    \item Why did you decide to participate in this study?
    \item How do you incorporate AI technology into your daily life?
\end{enumerate}

\subsection*{Establish Baseline}
Script: So AI-supported text technology has become extremely popular over the past year
Today we are going to center our discussion around two AI-supported text technology groups, AI text generators and autocorrect/ spell checkers. AI text generators use advanced natural language processing techniques to analyze existing text and generate new text that is similar in style and content, like your typical chatbots, smart text assistants, and chatGPT to name a few.

\begin{enumerate}
    \item Have you used or come across AI text generators, such as chatbots, smart text, assistants and ChatGPT?
    \begin{enumerate}
        \item \textbf{If yes:} Tell us about your time using AI text generators. 
        \item \textbf{Ask if needed:} Which ones are you familiar with? How often do you use them? What did you use them for? 
        \item \textbf{If no:} Move on to next question\\

    \end{enumerate}
\end{enumerate}

Script: Autocorrect, grammar and spell check are text editing features that identify misspelled words, and uses algorithms to identify the words most likely to have been intended, and edits the text, like on your iPhone, Microsoft Word or even Google Docs.

\begin{enumerate}
    \item Have you used or come across autocorrect, grammar or spell check like on your iPhone, Microsoft Word or even Google Docs?
    \begin{enumerate}
        \item  \textbf{If yes:} Tell us about your time using either of these.
        \item  \textbf{If needed:} Which ones are you familiar with? How often do you use them? What did you use them for?
        \item \textbf{If no:} Move on to next section. If they have said no to both of these questions end interview here. Be sure to thank them for their time. 
    \end{enumerate}

\end{enumerate}

\subsection*{Black Lived Experience}
Script: Culture refers to the shared beliefs, values, customs, behaviors, and artifacts that characterize a group or society. It encompasses everything from language and religion to food, music, art, and social norms, and helps to shape how people view themselves and others.

\begin{enumerate}
    \item How would you describe the Black American culture to someone who is not familiar with it... Imagine I was from outer space, and you were the first person I met and I asked you, “Tell me about the Black American culture”, what would you say to me?
    \item  In what ways do you think Black American culture differs from other cultures?
    \item If someone who was not Black could walk a day in your shoes as a Black individual, how might their experience differ?
    \item How would you describe the positive aspects of being Black in America? 
    \item How would you describe the negative aspects of being Black in America?
    \item What has your experience been like as a Black person in America?

\end{enumerate}

\subsection{AI-Supported Text Technology}
Script: We want to now transition into understanding how your experience as a Black individual in America shows up while using digital technology.

\subsubsection{Connecting the Black Experience to Technology}
\begin{enumerate}
    \item Do you see positive or negative aspects of your experience as Black person in America in your interactions with digital technologies?
        \begin{enumerate}
            \item \textbf{If yes:} What are the technologies? How?
            \item \textbf{If no:} What makes you say that?
        \end{enumerate}
    \item What aspects of your Black American culture do you NOT see in your interactions with digital technologies?
        \begin{enumerate}
            \item \textbf{Follow up: }What makes you say that?
        \end{enumerate}
    \item Has digital technology helped improve your experience as a Black individual?
        \begin{enumerate}
            \item \textbf{If yes:} How?
            \item \textbf{If no:} What makes you say that?
        \end{enumerate}
    \item Has digital technology worsened your experience as a Black individual?
        \begin{enumerate}
            \item \textbf{If yes:} How?
            \item \textbf{If no:} What makes you say that?
        \end{enumerate}
    \item Have you ever felt that technology was created specifically to address the needs or challenges faced by Black individuals?
        \begin{enumerate}
            \item \textbf{If yes:} How?
            \item \textbf{If no:} What makes you say that?
        \end{enumerate}
    \item Can you think of a time when you felt that digital technology was designed not having Black individuals in mind?
        \begin{enumerate}
            \item \textbf{If yes:} When?
            \item \textbf{If no:} What makes you say that?
        \end{enumerate}
\end{enumerate}

\subsection{Autocorrect/Grammar and Spell Check focused questions}
\begin{enumerate}
    \item What are your thoughts on autocorrect/ grammar and spell check?
    \item Do you think how Black people communicate through text was considered when autocorrect/ grammar and spell check was developed?
        \begin{enumerate}
            \item \textbf{Follow up:} Why or why not?
        \end{enumerate}
    \item Are there any features of autocorrect/ grammar and spell check that are exclusionary to Black people?
    \item If there were opportunities to improve autocorrect/ grammar and spell check to be more inclusive of Black language and culture, would you have any suggestions for what changes could be made?
        \begin{enumerate}
            \item \textbf{If yes:} What are they?
            \item \textbf{If no:} What makes you say that?
        \end{enumerate}
    \item Can you identify any ways in which autocorrect/ grammar and spell check addresses challenges faced by the Black community?
        \begin{enumerate}
            \item \textbf{If yes:} What are they?
            \item \textbf{If no:} What makes you say that?    
        \end{enumerate}
    \item Can you identify any ways in which autocorrect/ grammar and spell check is destructive or serve as a hindrance to the Black community?
        \begin{enumerate}
            \item \textbf{If yes:} What are they?
            \item \textbf{If no:} What makes you say that?    
        \end{enumerate}
\end{enumerate}
\subsection{AI-Text Generator focus questions}
\begin{enumerate}
    \item What are your thoughts on AI-text generators, such as smart text assistants, chatbots and chatGPT?
    \item Do you think how Black people communicate through text was considered when AI-text generators were developed?
       \begin{enumerate}
            \item \textbf{Follow up:} Why or why not?
        \end{enumerate}
    \item Are there any features of AI-text generators that are exclusionary to Black people?
    \item If there were opportunities to improve AI-text generators to be more inclusive of Black language and culture, would you have any suggestions for what changes could be made?
        \begin{enumerate}
            \item \textbf{If yes:} What are they?
            \item \textbf{If no:} What makes you say that?    
        \end{enumerate}
    \item Can you identify any ways in which AI-text generators address challenges faced by the Black community?
        \begin{enumerate}
            \item \textbf{If yes:} What are they?
            \item \textbf{If no:} What makes you say that?    
        \end{enumerate}
    \item Can you identify any ways in which AI-text generators are destructive or serve as a hindrance to the Black community?
        \begin{enumerate}
            \item \textbf{If yes:} What are they?
            \item \textbf{If no:} What makes you say that?    
        \end{enumerate}
\end{enumerate}

\subsection{Direct Observation of Technology Use} 

Script: We are now going to transition to the second portion of our interview. To begin you will need to access the link that I have just pasted in the chat. 
Pretend there is a time that you heard an interesting rumor/ gossip/ tea and you just had to text your bestie/ best friend. 
In at least 5 lines, we would like you to type out the story as if you were texting them now. 
Try to be as natural as possible in your writing, feel free to use slang or terms that you are most comfortable with.We are not here to test you but more so the technology that you are interacting with. Don’t worry about your grammar, spelling or anything of that sort. If you make a mistake, don't change or alter it.\\ 

*Give them two minutes to review and one minute to write, three minutes total*\\

Okay you may stop now. Let's go ahead and see what you put together. *Share your screen with the Google Doc visible*

 \begin{enumerate}
     \item What are your thoughts on the suggestions from Google docs? 
     \item Do you feel the suggestions from Google doc reflect your voice?
        \begin{enumerate}
            \item \textbf{If yes:} How?
            \item \textbf{If no:} Why is that?
        \end{enumerate}
    \item Are there any frictions with your natural style of text communication and the suggestions from Google doc?
        \begin{enumerate}
            \item \textbf{If yes:} How?
            \item \textbf{If no:} Why is that?
        \end{enumerate}
\end{enumerate}

Script: We are going to see how chatGPT takes your story and continues it. I will copy and paste your writing into the input box and we will discuss what it comes out with.

\begin{enumerate}
    \item How do you expect chatGPT to handle the rest of your story in regard to content and style of writing?\\
\end{enumerate}

Instructions for interviewer: *In chatGPT, copy and paste the following*: Continue my story with an additional ten more sentences ensuring to keep my tone and vernacular consistent: \textit{(insert the participant’s writing)}

\begin{enumerate}
    \item What are your thoughts on chatGPTs continuation of your story?
    \item Is the content of the story similar to something you would come up with?
        \begin{enumerate}
            \item \textbf{If yes:} How?
            \item \textbf{If no:} Why is that? What is missing? 
        \end{enumerate}
    \item Is the style of writing similar to yours?
        \begin{enumerate}
            \item \textbf{If yes:} How?
            \item \textbf{If no:} Why is that? What is missing? 
        \end{enumerate}
    \item In what ways do you think what chatGPT wrote represents or mis-represents your identity as a Black individual?
    \item Some people see chatGPTs output in American English and not African American English as an issue, what are your thoughts?
\end{enumerate}

Script: These are all of the questions that I have for you today. I really enjoyed hearing your thoughts and stories surrounding your experiences with AI-supported text technology. Our team sincerely thanks you for taking part in this study and disclosing such personal information to us. You will be hearing from us by early May for the next portion of the study.\\ 
Before I let you go, do you have any other thoughts or feedback on your experience participating in this study?\\
Would you like a copy of the recording?\\
Thank you for your time with us, we hope that you have a great rest of your day!
\newpage

\section{Participant Overview}

\begin{longtable} {p{6em} p{30em}}
\caption{Overview of individual participants' usage with AISWT}\\
 \toprule
    Participant ID & Context for individual AISWT usage \\
    \midrule
    \endfirsthead
    
    \toprule
    Participant ID & Context for individual AISWT usage \\
    \midrule
    \endhead

    \texttt P1 & Engages with autocorrect, Grammarly and ChatGPT for tasks ranging from creating emails to writing documents. Their encounters extend to ChatGPT in an educational setting for class exercises, experimenting with diverse prompts to analyze responses. \\
    \texttt P2 & Leverages ChatGPT as a versatile writing companion for various tasks, including crafting essays, planning ideas, and generating travel itineraries. Frequents Grammarly for spelling, grammar, clarity, and flow, and utilizes autocorrect on their phone as a consistent part of their daily writing routine.\\
    \texttt P3 & The individual frequently utilizes ChatGPT employing it two to three times a week. They consistently rely on autocorrect and spellcheckers in their regular writing routine. \\
    \texttt P4 & Relies on ChatGPT for problem-solving and generating baseline code in their professional and academic endeavors. Also explores playful interactions and tracks daily calorie intake using ChatGPT on a personal level, while autocorrect, spellcheckers, and Grammarly play distinct roles in their daily writing routine. \\ 
    \texttt P5 & Relies heavily on ChatGPT for automating daily tasks, utilizing it extensively for formatting emails, improving text structure, and refining grammar in various contexts, including answering emails and crafting recommendation letters. Everyday texting benefits from autocorrect and autopredict features. \\
    \texttt P6 & Employs ChatGPT for personal projects and communication, such as structuring a script and storyboarding for short films, and uses autocorrect and spellcheck daily. \\
    \texttt P7 & Frequently relies on chatbots for online customer service interactions. Additionally, they employ autocorrect and spellcheck across platforms like Google Docs and Microsoft Word.\\
    \texttt P8 & Actively engages with ChatGPT and utilizes chatbots for shopping assistance. In addition to Grammarly, they leverage Notion's AI capabilities to enhance language clarity and tone in written communication, and they are familiar with autocorrect and grammar features on platforms like Android, Microsoft Word, and Google Docs. \\
    \texttt P9 & Uses ChatGPT to enhance the quality of their written communication, using it for crafting polished emails and essays to present themselves as a better student. They specifically utilize ChatGPT for paraphrasing and rely on autocorrect, primarily on Microsoft Word and Google Docs, to improve the overall clarity and conciseness of their written content.  \\
    \texttt P10 & Uses ChatGPT, for both personal and work-related tasks, leveraging it for tasks ranging from generating JavaScript code and chatbots to enhancing responses on Teams. Regularly employs autocorrect and spell check features, particularly in email correspondence through platforms like Outlook and Gmail, emphasizing the context of their usage in improving written communication and work-related tasks.\\
    \texttt P11 & Relies on ChatGPT and Quillbot for academic assignments, seeking clarity and precision in their responses. In addition, they specifically use Gboard, Google's autocorrect tool, to enhance text accuracy, emphasizing the academic context of their usage. \\
    \texttt P12 & Utilizes chatbots for work communication relying on virtual assistants for note-taking during meetings. Leverages Word AI for concise sentence structuring in professional communication and heavily depends on autocorrect, grammar, and spellcheck features in their iPhone and Microsoft Word for personal text-related tasks. \\
    \texttt P13 & Relies heavily on autocorrect and spellcheckers for phone and Word typing. While having experience with library and banking chatbots, their occasional use of ChatGPT is specific to academic needs, such as designing lesson plans for classes. \\
    \bottomrule
\end{longtable}
\newpage

\section{Codebook}

\begin{longtable}{p{7em} p{12em} p{18em}}
\caption{Codebook generated through analysis}\\
\toprule
                 Theme &                                         Definition &                           Example from Transcripts \\
\midrule
\endfirsthead

\toprule
                 Theme &                                         Definition &                           Example from Transcripts \\
\midrule
\endhead
          Afterthought & Incorporating Black culture into design only after public backlash or as a performative gesture for recognition and praise. & "Oh, man, when it comes to like a research just in general, especially when it comes to like data science and technology. I feel like they don't take into account, you know, different, like racism, or other different things. So I saw a couple of cases, like where, you know, Apple with the facial recognition software. You know, I have a friend who works there. And they said that they needed they started like talking like black people from the workplace and scanning the face to try to, like add the facial data to like some of that stuff. And then I see some of the biases and AI and data and I'm like, okay, like there needs to be more black research just to help, you know, shape and determine outcomes with AI stuff and alone. So that's why I want to participate." - Black Tiger \\
        Black Identity & How the user's identity shapes their experiences and influences the way they navigate the world. & "I would say to you that this is my culture and in my culture we are the resilient ones because we are the ones that get even they get hit most times but then we still stay at the top form. We are simple people and then we encourage simplicity.... someone who's no black walks in my own gig the experience I think they will be super amazed at the kind of strength they have now it would be different because as a black person, you just have to be strong so you always got to have the strength" - P3  \\
         Black Support & How the user's identity shapes their experiences and influences the way they navigate the world. & "So whether that's joining blacksmith Association, or Ethiopian Student Association, things like that definitely helped me build community and find home, and then post grad, post post undergrad, I started working and was able to find communities, again, that were supporting, you know, black resource groups at work. And I had the opportunity to be very intentional about like, the work that I was doing, and to sort of be able to give back, and now kind of hopping back into academia" - P1 \\
         Design Reflection & Users' thoughts on how system design impacts their experience. & ``You're gonna laugh, because the only example that comes to mind is hinge. So, you know, I'm on the apps swiping, you know, doing what I do, not a fan of swiping, but here we are in 2023. And one thing I started noticing I told some, my girlfriend's like, you guys, just correct me if I'm crazy. But this is something I've noticed as a trend is, obviously there are algorithms at play, and they're seeing what kind of individuals you're swiping on, I'm sure they're taking the demographic information and like plotting things and finding folks who are similar to, you know, to display to you. And there'll be days, and I'm like Colorado's, diverse enough, but it's not the most woke slash black friendly place. I remember swiping, and I had like, 50 black men, like one after the next. And then another day had all Asian men, one after the next. And then the next time I had all Caucasian men when I was like, this is weird. This is actually crazy thinking that it'd be like a random sample, like a random bag. I get when it is when that, you know, no, literally one after the next for. Yeah, decades of profiles. It was very strange, and it still happens today.'' - MamaAfrika \\
   Design \\Requirements & Key elements that should be incorporated into the product to make it ideal. & "Yeah, well, um, Google and Apple. They're definitely trying, I noticed, you know, they had the skin tones change. But I still feel like they could expand. I mean, something as small as like expanding the type of emoji they offered. Just adding things that are from our culture in there, I think would be nice. I don't know. Yeah. And then maybe also with the filters, the face filters, just, yeah, it seems like they aren't tested with people who have our features. And skin tone maybe." - Blue Bird \\
                Equity & All users, regardless of race, can access technology with the same ease and capabilities. & "I think it's going to have all the Black people to be able to confidently communicate with folks from all around the world and in such a way that its correct and engaging." - P3 \\
   Exclusionary Design & Design that neglects to account for the specific needs and use cases of Black users. & "So I was wearing like a head wrap that had like a button in the back. And like the head machine wasn't wide enough to go past it, so kept knocking me in the head. And like, eventually, like, you know, we made it work. But the dentist had told me after like, yeah, like, they did not design this machine to really consider different types of like, things that people wear on their heads. Because because she had another she had another, had another black man who had like his head, he had like his, his he had really long locks, and they were rolled up into a bun as well. And that also was like hitting him when it was going around too. So that's what that's what I think about. I think another thing is like, we think we go to TSA. I always feel like it's black women who have to have their hair touched. Um, maybe that's not the case." - P2 \\
  Exclusionary Editing & AI editors often recommend editing or removing words and names commonly used by Black users. & "I mean, I kind of understood it, because that's a parameter they said on it. It was like, oh, we can't write anything with offensive language or something like that. I forgot to you know, to spell did it gave me but, um, it showed that it does have parameters, and it can, you know, be controlled, essentially, but, you know, that word is a part of our culture. And, you know, I couldn't you know, write a script or write out that part of the script. I had to write that part of script myself because the word had to be implemented into that space. Like, there's no way I could, because, you know, instead of like five or six times when I was writing it out, when I entered it in there, I only use it once and like, wouldn't touch it. So but, I mean, maybe I was kind of happy in a way because it was like no, but at the same time like you You can still get it to do what you want to do, you just have to change the parameters on it. So if I would have, you know, put in dot, dot, you know, whatever and kind of spelled it out like that, then if they would have put that word in there, but I wouldn't have known it that way. It just depends." - Black Tiger \\
  Highs of being Black & Positive and empowering aspects of being Black in America. & "Um, I think black is beautiful. Like, I think black is creative, being black is is to be creative. It's to be innovative, resourceful. I like those are some of the, I guess, like, key adjectives that come to mind? I think resourceful is an interesting one. Because like I mentioned earlier, like, obviously, historically, there have been you know, this country, the United States hasn't served black Americans. And so I think that idea of being resourceful comes from a place of of hurt and pain, but has led to like innovation and creativity and you know, things that are beautiful, right? So I think yeah, like that. Those are some of the adjectives that that come to mind when describing the black experience in the United States." - Purple Lizzard \\
      Inclusive Design & Technology is designed with inclusivity at its core, ensuring global relevance and equity. & "Um, I wonder if like, if, if in a prompt you you're using language that is commonly using black communities, if, if the model is not familiar with how to respond or is, I guess, not certain, or I don't know how you would even determine certainty here but let's just let's just say like that was already predetermined. Then some sort of like response, say saying, like, based on your prompt, this is what I've understood. And based on what I've understood, this is my answer to your question. I think that might be helpful. I also think in the future would be really interesting to see the model respond in that same language. I think that would be very interesting. Um, yeah, but I just I'm not exactly sure how that would work." - Purple Lizzard \\
Internalized Construct & Exploring how marginalized individuals may internalize discriminatory beliefs. & "I always feel like you're being scrutinized just for like breathing. Right? Yeah. Um, I really feel like that's it, that's really just like, again, go at having to be on going back to being on guard, right? That's just exhausting. It's mentally taxing. Um, always, like, feel like granted either, like, not all black people go through this, but it just like, for me, it just like always questioning myself, right? Like, did I do this wrong? Or am I am I am I able to do this? Right? Do I have enough experience? Like, again, like, the I think the psychological gymnastics you have to do to, to really just like survive is shitty. And even then, like also being being in a place where you get so much knowledge and know so much about, like how the world works, and especially like how the US works is also very, you deal with a lot of anger, too. So it's also been something as well, I think it was who said that? It's like, something along those lines, like, the more educated you get about like the systems that work particularly in the US, James Baldwin?" - P2 \\
   Lows of being Black & The challenges and hardships of being Black in America. & "I'm sure you already know as a black man, right? It's hard. It's hard. being black in America and it's something that having kind of grown up here. A lot of my cousins back home don't understand. So they assume that you know, America's the land of milk and honey money just grows on trees, everyone's Kumbaya, but I have a cousin who is now attending university here in Florida. And he had a rude awakening. Right? He was in DeLand, Florida, when a very few black folks at the time I think now the school has increased their diversity. But it was it was a shock to him, right of all the things I told him, I'm worried, I'm like, Hey, you should read up on American history. Because these are things that you don't think about here in Ghana, that you're going to unfortunately, as a black man in this country, given the history, especially what we're living through right now, this is a very real possibility for you, and I don't want you to be caught off guard. So I think that the black experience is one that you have to kind of tread cautiously. And that's a really unfortunate, because I feel like in some places, you can't fully be black, I think there's an expectation to assimilate to white culture, right. And I'm guilty of it, right. straightening your hair and skin bleaching and all these things that are really unfortunate fear of authority figures. There's just a lot of things that come with being black that I think other cultures and other folks maybe don't necessarily have to deal with on a daily basis." - MamaAfrika \\
        Mixed Emotions & A mix of positive and negative emotions related to participants' experiences. & "So there were parts that I was like, okay, like, I see where you're going. And I could see how like the tone was trying to be there. But I do think that like, the third paragraph there is very serious. Like, I don't know how it became this coffee became like, that's not how the real world works. Like, I don't I don't know how I got that serious. Like, you have to be responsible for your actions, I think is definitely a heavy statement. for spilled coffee. In my personal opinion. So yeah, so I do think that it kind of, like faded, I guess in terms of like that the the tone that I was trying to use in my original message. Yeah, it also feels like a script. Like, I don't know, like, it doesn't feel like I like someone would actually say this, like, even via text or even phone call. Like, I'm not exactly sure if I would say you know, they think they can do whatever they want. Like I would see like, I would even like rephrase that to be like, like man like people are like I would say like think they can do whatever you know, whatever you want. Like not gonna get caught up for example, which is like really the same thing as saying there won't be any consequences but I'm just not sure if I would say that in this in this situation. So definitely opportunity to sort of change the tone or to match the original text here, original prompt so." - Purple Lizzard \\
     Negative Emotions & Negative emotions triggered by participants' experiences. & "Sometimes it makes me feel like I'm kind of dumb is dumb. Yeah, I'll say dumb. Like, my English is not the best. But I do know that my English is the best. It's just sometimes it's it's a bit different, or it has more embellishments in there." - P9 \\
     Positive Emotions & Positive emotions evoked by participants' experiences. & "It's great for me. I like it. I use it all the time. And for, for having to write a paper where it's like, it has to be this way, and there's no wiggle room for it. I'm all about it. So the only thing about spellcheck and this was also mentioned on social media specific words, like well, AutoCorrect, when you're trying to, like you're trying to say a specific word in a community. That means something that might not be nice to say, but autocorrect to something else. And so I noticed that that was a popular, a popular topic that was trending. And they were talking about that, like on the news and stuff like that. That's the only time and that rarely happens. That's the only time where I don't see it working out. But I use it all the time. And it's really convenient. I have no issues with it." - P13 \\
  Psychological Impact & Examining the emotional, psychological, and mental health impacts from discrimination. & "Are they people are treating me different because of like how my hair looks or, or my phenotype as, granted I'm light skinned, but I do have more of like black, you know, like a black more of a black phenotype. So just like, if someone were to treat me different or even like if someone says something snarky to me, like you always have to question was that because I am black or because I'm a black woman, right? It's Oh, you always have to think about these things. And it sucks. Though, we have to be on guard. That's what it feels like always to be on guard always be like out there protecting myself, especially when I feel like I always have to advocate for myself, especially like being this whole graduate program. I always feel like I have to fight for my life. And always support myself because it feels like no one else has really done that except me and other black woman. So yeah, so like when it comes to this navigating, where I feel like I'm always on guard. And like, even though you even notice the wrongdoings and stuff, too, it's just like, and so you feel like you're the one the only one who speaks up, or something's wrong. Right?" - P2 \\
                Racism & Instances of discrimination and judgment participants experienced based on their identity. & "Are there even biases and how like, the like, how the, you know, the filtering systems that they use to go through resumes and stuff? Or like, are they gonna see, I mean, we already saw it like, basic, like names and stuff, I'm not sure if like, that's like in the AI systems, I know that that's like at the human level. But if you're training again, if you're training these AI systems, and you're biased yourself, right, is going to be in, like in the system, right? So like, you've ever think about like, Oh, if this person doesn't have like, a white sounding name, right, then that's, that's it? No, or like, it can also be used dangerously to like, Oh, they're involved in a lot of like, like, woman of like, like, people of color centric things, right? Like, oh, I'm part of like the National Black honor society Who knows, right, like, so it's just like, that's really, that's really really scary. Because I really feel like I can definitely use it a — is being used it can be continued to use to exclude us. I'm terribly I'm really trying to get like other ways that it's done." - P2 \\
               Reliant & A growing dependency on AI, contributing to a decline in critical thinking and independence. & "It's made me lazy. It's kind of like the calculator, right? Like um I don't know how to spell anymore. Um because we're so reliant on these technologies of this technology to do it for us. Because like someone had just, this is so funny. I was just in a, in a call where one of the activities was like to spell camouflage. And like I'm sitting here, I was like, damn, I don't know how to spell camouflage. And like, you see all these other people put in the chat. Like they spelled it correctly and stuff. And then like, and then like, the presenter was like, Wow, you guys really know how to spell camouflage. That's really, really good. It's like the first time like, when, like when the majority of people knew how to spell it, right. And then someone said, like, yeah, just just autocorrect! Because you can, it's on Zoom. Right. So, so, um, so yeah. I just like, it's anyway, just to go to say like, like, not camouflage. Excuse me. Um, yeah, autocorrect has really made me lazy. And also, I feel like I'm really relying on it to spell things out for me, especially like those really tricky words in English. So you just like, you wouldn't think it's spelled that way, but it is spelled that way. Um, so most definitely, yeah." - P2 \\
      Style Conforming & Participants used automated editing to achieve "professionalism" in sentence structure. & "I will say like, I have like a because you know, ChatGPT I look at it as a way to just to ease to I look at it as a platform that you can communicate with the computer a lot easier. So I created a part I say, ChatGPT create an email about create email that introduces me to this person, I create an email with this particular subject topic I put the prompt in. And then you know, I have a few more parameters, different things that I use to kind of like match the email but I do like to tone the Email, to style the email, you know, obviously, you know, either the first person, third person or what I want to say and kind of do it that way. But it depends, like, if I'm emailing a friend, I tell it, you know, this is a friend, this is the language I want you to use, or this is a professional and this is what I want you to use, and kind of do it that way. So I frame it, you know, I put different parameters and and depending on who I'm talking about what the message would be how long I want it to be." - Black Tiger \\
     Style Consistency & Users perceive that AI-generated text does not impact their personal writing style. & "I think it does didn't really correct anything? Maybe because I was taking too long to think about it. With some juicy gossip that I have, um Yeah, it didn't I was surprised about that thing about baby mama. Yeah, yeah, I was. I was surprised about that. Um, it didn't it didn't capitalize January. Yeah, yeah. It's like it pretty. It doesn't affect my voice." - P2 \\
      Style Dissonance & Users feel that AI-generated text does not reflect their personal voice or style. & "It's cute. I think it's a nice like, little narration but it's not like how I would go about it. Yeah, it's, you know, I mean, I think it's being repetitive too I feel like it's just saying the same thing." - P2 \\
              Task Aid & The process of generating ideas and tasks like outlines and itineraries with AI support. & "So I use ChatGPT was one thing that recently came out. And I use it to automate certain tasks that I do on a daily like when it comes to formatting emails, learning formatting text, say for example, I have some paper I would use it to format to prediction in the grammars they're looking for and I would use ChatGPT to format the text the format the text by asking to fix grammar, diction. From typing after better ways, more more correct ways of saying things. When it comes to email, sometimes I need to quickly respond to something, I would ask it, for me, the template does something for me to, to use and to format my, my response in that way. So I think those are the things that I use it for this reason, mainly for like automating certain tasks that I do done previously, such as like reading texts or answering emails or recommendation letters difference." - P5 \\
      Underrepresented  & Aspects of identity and personal style outside societal norms can lead to exclusion. & ``Yeah, I'm, I think, like, definitely with like spellcheck. I'm like some words that are used in the way that like, like language that's used by black people may be considered spelled like incorrectly Um, by like, I guess, you know, whatever application that you're using, so I definitely think that could be considered exclusionary, right. Because it's like, it's incorrect to who and like, you know, it kind of brings that question of like, what is formal? Like, what is correct? What is to be considered? a correct way of speaking? So yeah, no, I definitely think that there, there are aspects that are very much exclusionary.'' - P1 \\
\bottomrule
\end{longtable}

\end{document}